# Sulfur in the Moon and Mercury

**Christian J. Renggli[1,2], Edgar S. Steenstra[2,3], Alberto E. Saal[4]**

[1] Max Planck Institute for Solar System Research, Göttingen, 37077, Germany
[2] Institute for Mineralogy, University of Münster, Münster, 48149, Göttingen
[3] Faculty of Aerospace Engineering, TU Delft, Delft, 2629, The Netherlands
[4] Department of Earth, Environmental & Planetary Sciences, Brown University, Providence RI, 02912, USA
ORCID: Renggli 0000-0001-8913-4176

## Abstract

This chapter presents a comprehensive overview of the abundances and distribution of S, and the processes that control the behavior of S on the Earth's Moon and on Mercury. The two planetary bodies share notable similarities, such as lacking substantial atmospheres and featuring surfaces with high numbers of impact craters. Both objects are at variably low oxygen fugacities ($f$O$_2$), where S occurs only in its reduced state as S$^{2-}$, and forms sulfides. For the Moon, we present a compilation of 55 years of lunar sample analysis, from Apollo 11 to Chang'e 5, including S concentrations and isotopic compositional data. We discuss processes from S in the lunar interior to volcanic degassing from mare basalts and in pyroclastic eruptions. At the beginning of a lunar science renaissance, we highlight where future research into S on the Moon might lead. The knowledge of S on Mercury is almost entirely based on observations by the NASA MESSENGER mission, operational from 2011 to 2014. MESSENGER observed very high S abundances on Mercury with concentrations of up to 4 wt.%. We discuss proposed mechanisms that lead to the high S abundance on Mercury's surface and discuss implications for the planet's interior. Finally, we provide an outlook on how the upcoming ESA/JAXA BepiColombo will advance our understanding of Mercury and the processes controlling S on the planet.

# 1. Introduction

A comparison of the Earth's Moon and Mercury reveals remarkable similarities, yet their geologic histories, controlling the inventory and behavior of sulfur, differ considerably between these planets. First, both lack substantial atmospheres. Second, the surfaces of these planets are characterized by high numbers of impact craters, suggesting limited magmatic and geologic activity over the last few billion years, recording processes that mostly occurred more than 3 Ga ago (Rossi et al., 2018). Third, both the Moon and Mercury are very reduced, at oxygen fugacities where S only occurs as reduced sulfide (Wadhwa, 2008; Zolotov, 2011). Mercury is larger with a radius of 2438 km, compared to the Moon with a radius of 1737.4 km, and Mercury has a very high density of 5.43 g/cm$^3$, compared to the Moon's density of 3.34 g/cm$^3$ (Spohn, 2015; Andrews-Hanna et al., 2023). In the introduction to this chapter, we provide a brief overview of these planets and discuss the behavior of S at reduced conditions generally, before we present the S record for the Moon and Mercury in detail.

The extent, timing, and composition of lunar magmatism are key pieces of information to understand the origin and evolution of the Moon, including the behavior of volatiles. After the Moon-forming giant impact event, the cooling and crystallization of the primordial lunar magma ocean (LMO) created an anorthosite crust, layered igneous cumulates, and late-stage components enriched in Ti and incompatible elements (ilmenite-rich layer and urKREEP) (Snyder et al., 1992; Elkins-Tanton et al., 2011; Dygert et al., 2017; Schmidt and Kraettli, 2022). Rayleigh-Taylor instabilities or a full overturn of the cumulate pile (Hess and Parmentier, 1995) may have caused 1) the sinking of the late-stage dense layer that mixed with the earlier cumulates, and 2) the rise and melting of the early LMO cumulates to the base of the primordial crust, generating the Mg-suite of plutonic highland rocks <10 Ma after the primordial differentiation of the Moon. Thus, the LMO crystallization and cumulate overturn is believed to be the cause for the heterogeneous source region responsible for the geochemically diverse anorthosite crust, Mg-suite plutonic rocks and lunar magmatism, from KREEP (K-, Rare Earth Element and P-rich) to mare basalts (Shearer et al., 2006; Shearer et al., 2023) with volcanism over a period of ~2.7 Ga, from 3.93 Ga to 1.2 Ga (Hiesinger et al., 2003, 2023; Head et al., 2023). The lunar core makes up less than 2% of the Moon's mass. A recent reexamination of Apollo era seismic data indicates the presence of a solid inner core with a radius of 258 ± 40 km, an outer core with a radius of 362 ± 15 km, a low velocity zone at 560 ± 34 km, the top of the mantle at 1698.6 km radius, and a crust with a thickness of ~40 km, resulting in an overall radius of 1737.1 km (Briaud et al., 2023).

The magmatic evolution of Mercury is less constrained compared to the Moon. However, volcanic activity is recorded in basaltic smooth plains, with ages as young as 1.5 Ga (Denevi et al., 2018). The silicate mineralogy at the surface of Mercury consists primarily of minerals common in basaltic rocks. The major rock-forming minerals predicted from experiments and normative calculations are plagioclase (albite to anorthite$_{50}$), pyroxene (diopside and hypersthene), and olivine, which places the majority of the investigated terranes mineralogically in the olivine gabbro, or olivine norite fields (Namur and Charlier, 2017; Vander Kaaden et al., 2017; McCoy et al., 2018). Mercury has a large metallic core (approximately 65% by mass) and only a thin silicate mantle, from which a relatively thick crust formed by partial melting. Mercury's radius is 2440 km, with a core-mantle boundary at 420 ± 30 km depth (Hauck et al., 2013), and an average crustal thickness of 35 ± 18 km (Padovan et al., 2015). See McCubbin & Anzures (2024) for a recent review on the chemistry and petrology of Mercury in general.

Detailed observations of the Moon go back to the first map by Galilei published in the Sidereus Nuncius (Galilei, 1610), where he identified abundant craters. The understanding of the Moon has vastly changed with the US Apollo program, which returned 382 kg of lunar samples, and the Soviet robotic Luna missions that returned 326 g of lunar samples. Recently, the Chinese probe Chang'e 5 returned with 1.731 kg of lunar materials (Che et al., 2021; Hu et al., 2021; Li et al., 2021, 2022). At the time of completing this chapter, Chang'e 6 has returned with 1.9353 kg of material, the first samples from the lunar farside, however S abundances in the samples have not yet been reported (Li et al. 2024). Additionally, lunar materials have been recognized in meteorite collections, totaling to more than 60 kg (see https://curator.jsc.nasa.gov/antmet/lmc/lunar_meteorites.cfm). Since the late 1990s lunar science has experienced a revival, driven by advances in analytical techniques and (re)-investigations of the legacy samples from the Apollo era, as well as a suite of remote sensing missions, including Clementine, Lunar Prospector, GRAIL and Lunar Reconnaissance Orbiter from the US, Hiten and SELENE from Japan, Chandrayaan-1 and -2 from India, and several Chang'e missions from China (Neal, 2009). As several nations prepare for new sample return and crewed missions to the surface of the Moon, we can expect a growing suite of new lunar samples over the coming decades. In this chapter we present a compilation of measured S concentrations and S stable isotope compositions in lunar samples and discuss what has been learned about S from more than 50 years of lunar sample science.

In contrast, Mercury is the least investigated terrestrial planet in the inner solar system. The first detailed observation of Mercury occurred in 1974-1975 when the US probe Mariner 10 had three flybys of the planet, imaging ~45% of the surface (Strom, 1979). The first evidence for elevated S abundances on Mercury came from ground-based mid infrared thermal emission spectroscopy and suggested the presence of elemental S at the poles, and sulfides in the regolith (Sprague et al., 1995). The US probe MESSENGER was only the second mission to Mercury and orbited the planet from March 2011 to April 2015 (Solomon and Anderson, 2018), providing the only constraints on the planet's S inventory from remote X-ray fluorescence spectroscopy (XRS) and gamma ray spectroscopy (GRS). These methods allowed the measurement of S relative to Si (S/Si element ratios) from the top tens of micrometers of the surface (XRS) to some tens of centimeters into the surface regolith (GRS) (Nittler et al., 2018). The European and Japanese mission BepiColombo is the next Mercury probe and will enter the planet's orbit in December 2025, investigating the planet and its magnetic field, including chemical and mineralogical investigations by IR, UV, X-ray, γ-ray, and neutron spectrometry (Benkhoff et al., 2010; McNutt et al., 2018).

The oxygen fugacities of planetary interiors, and the magmatic and crustal rocks derived from them, are important for the understanding of the behavior of S during planetary processes due to the polyvalent nature of the element, with a valence range from 2- to 6+ (see also Franz, 2024 this volume; Harlov & Pokrovski, 2024 this volume; Komabayashi & Thompson, 2024 this volume; Lodders & Fegley, 2024 this volume; Schrader et al., 2024 this volume; Zolotov, 2024 this volume). The mantles of the Moon and Mercury are so reducing that sulfur likely only exists in the reduced state $S^{2-}$, in sulfides, silicate melts, or volcanic gases. The oxygen fugacity of the lunar mantle and erupted basalts is determined relative to the iron-wüstite buffer (IW) at IW+0.2 to IW-2.5 (Fogel and Rutherford, 1995;

Wadhwa, 2008). The lunar oxygen fugacity is based on a number of different direct and in-direct measurements, such as intrinsic oxygen fugacity measurements with a solid-electrolyte oxygen cell (Sato et al., 1973; Sato, 1976), mineral equilibria in lunar basalts (Haggerty and Meyer, 1970; Nash and Hausel, 1973; Delano, 1990; Steele et al., 1992), spectroscopically determined C abundances in pyroclastic glasses (Fogel and Rutherford, 1995), or the valence state of vanadium and Fe in pyroclastic glasses (Karner et al., 2006; McCanta et al., 2017).

For Mercury, the oxygen fugacity of erupted basalts and the mantle is less well constrained, but a number of observations indicate even more reducing conditions below IW-3. The primary evidence for the very low $fO_2$ comes from the low abundance of FeO (< 3 wt.%) on the surface that had been inferred from Earth based spectral observations (Robinson and Taylor, 2001; Warell and Blewett, 2004), and constrained by MESSENGER to 0.6-2.4 wt.% FeO (Nittler et al., 2011; Weider et al., 2016; McCoy et al., 2018). Together with an elevated abundance of S on the surface (a detailed discussion of S abundances comes later in this chapter), and under the assumption that S was dissolved in a silicate liquid, the $fO_2$ is estimated to IW-2.6 to IW-6.3 (Malavergne et al., 2010; Zolotov, 2011; McCubbin et al., 2012; Zolotov et al., 2013; Namur et al., 2016).

Here, we briefly recapitulate the behavior of S in silicate melts and volcanic gases at very reducing conditions relevant for the Moon and Mercury. Below ~IW+3, where S is dissolved as $S^{2-}$ in silicate melts (Jugo et al., 2010; Wilke et al., 2011; Simon et al., 2024 this volume), it replaces oxide anions ($O^{2-}$) on the anion sublattice, and the solubility of S increases with decreasing $fO_2$ (Fincham and Richardson, 1954; O'Neill and Mavrogenes, 2002; Wykes et al., 2015; Namur et al., 2016). The increase in S solubility with decreasing $fO_2$ is accompanied by a change in sulfide speciation in the melt, from FeS above IW-2, to CaS and $Na_2S$ at IW-3, and primarily MgS below IW-4.5, as shown in Figure 1 (Anzures et al., 2020b). The sulfide complexation in melts changes chemical and physical melt properties, including an increase of the $SiO_2$ activity, shifting the phase stability from forsterite to enstatite, and increasing the melt polymerization and viscosity (Holzheid and Grove, 2002; Namur et al., 2016; Anzures et al., 2020b; Mouser et al., 2021). The oxidation of a melt from IW-7 to IW-3 results in a decrease of the S solubility from ~7 wt.% to ~1 wt.% (Figure 1).

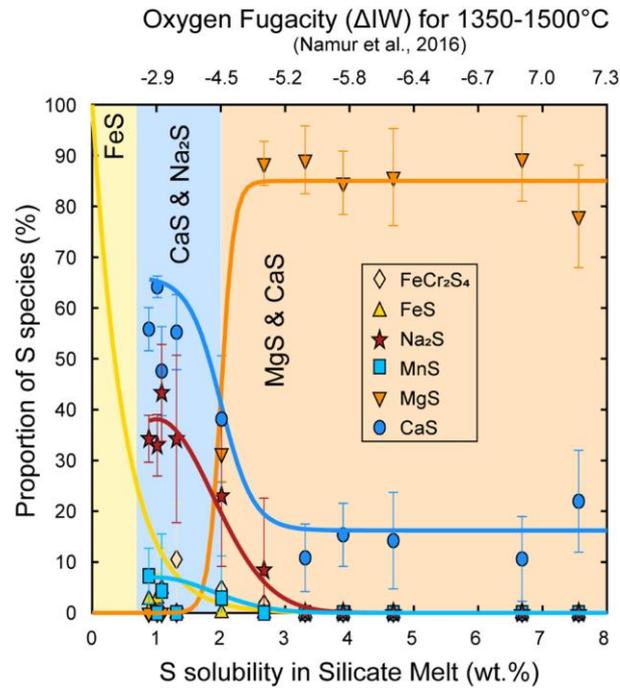

*Figure 1. Sulfur speciation in glasses determined by X-ray absorption near-edge spectroscopy (XANES) as a function of S concentration and oxygen fugacity relative to the IW-buffer (reproduced from Anzures et al. 2020b).*

In the gas phase S occurs in a diverse mix of species varying as a function of $fO_2$, with $SO_2$ dominant at oxidizing conditions above IW+3. At reducing conditions S gas species include $S_2$, COS, $CS_2$, $S_2O$, CS, and $S_3$ in water free systems (Figure 2a), and $H_2S$ in water bearing systems (Figure 2b) (see also Casas et al., 2024 this volume; Eldridge et al., 2024 this volume). The calculations shown in Figure 2 reflect the speciation of volcanic gases on the Moon and Mercury, with the exception that halides are not included (Fegley, 1991; Zolotov, 2011; Renggli et al., 2017). For the comparison we chose a simplified system with 1 mol% S (Figure 2a) and additionally 2 mol% H (Figure 2b) in a CO-$CO_2$ gas mixture. These compositions serve to illustrate the $fO_2$ dependent change in gas speciation and are not based on specific compositions for the Moon or Mercury. The molar ratios of S/C, H/C, and H/S are controlled by the volatile abundances in the magmatic sources, with S/C = 2.18, H/C = 2, and H/S = 0.9 on the Moon (Saal et al., 2008; Renggli et al., 2017), and are unknown for Mercury (Zolotov, 2011). In summary, likely the main difference between S gas speciation on the Moon and Mercury is the higher abundance of the species $CS_2$ and CS on Mercury (Figure 2).

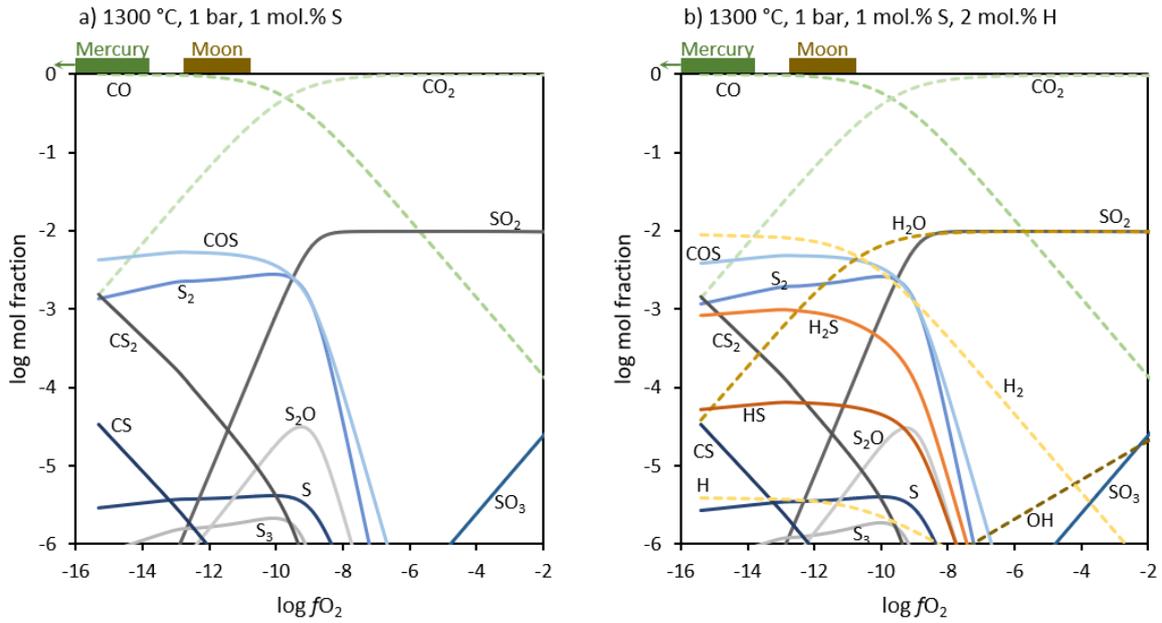

*Figure 2. Gas-phase speciation of S in a CO-CO$_2$ gas mixture at 1300 °C, 1 bar, as a function of log fO$_2$. a) S-C-O system. b) H-S-C-O system with a molar H/S = 2. Gas phase speciation models at 1300 °C and 1 bar as a function of fO$_2$, calculated by Gibbs free energy minimization with the software package HSC9 by Metso:Outotec. In the calculations fO$_2$ was varied by stepwise adding O$_2$ to an initial set composition of 99 mol% CO and 1 mol% S in the water free system (Figure 2a), and 97 mol% CO, 1 mol% S, and 2 mol% H in the water bearing system (Figure 2b). At 1300 °C and 1 bar the IW buffer is at log fO$_2$ = -10.75. The oxygen fugacities of the Moon (IW to IW-2, log fO$_2$ from -10.75 to -12.75) and Mercury (below IW-3, log fO$_2$ < -13.75) are marked on the figures in brown and green respectively.*

## 2. Moon

### The S inventory in lunar samples

In Figure 3, Figure 4, and Figure 5 we show a compilation of S concentrations and $\delta^{34}$S ($\delta^{34}$S = [($^{34}$S/$^{32}$S)$_{sample}$/($^{34}$S/$^{32}$S)$_{V-CDT}$ -1] × 1000, V-CDT is the Vienna-Canyon Diablo Troilite international standard, see Eldridge et al., 2024 this volume) stable isotope compositions determined for mare basalts, picritic glasses, impact melt breccias, regolith breccias, soil (regolith) samples, magnesium suite and anorthosite samples, and olivine hosted melt inclusions. These data were obtained by a number of different analytical techniques including sample combustion, acidification, SIMS (Secondary-Ion Mass Spectrometry), nanoSIMS (nanoscale SIMS), and EMPA (Electron Micro Probe Analyzer), from US Apollo missions and Soviet Luna missions. The complete dataset, including major element compositions and detailed references is available on a repository (Renggli, 2023).

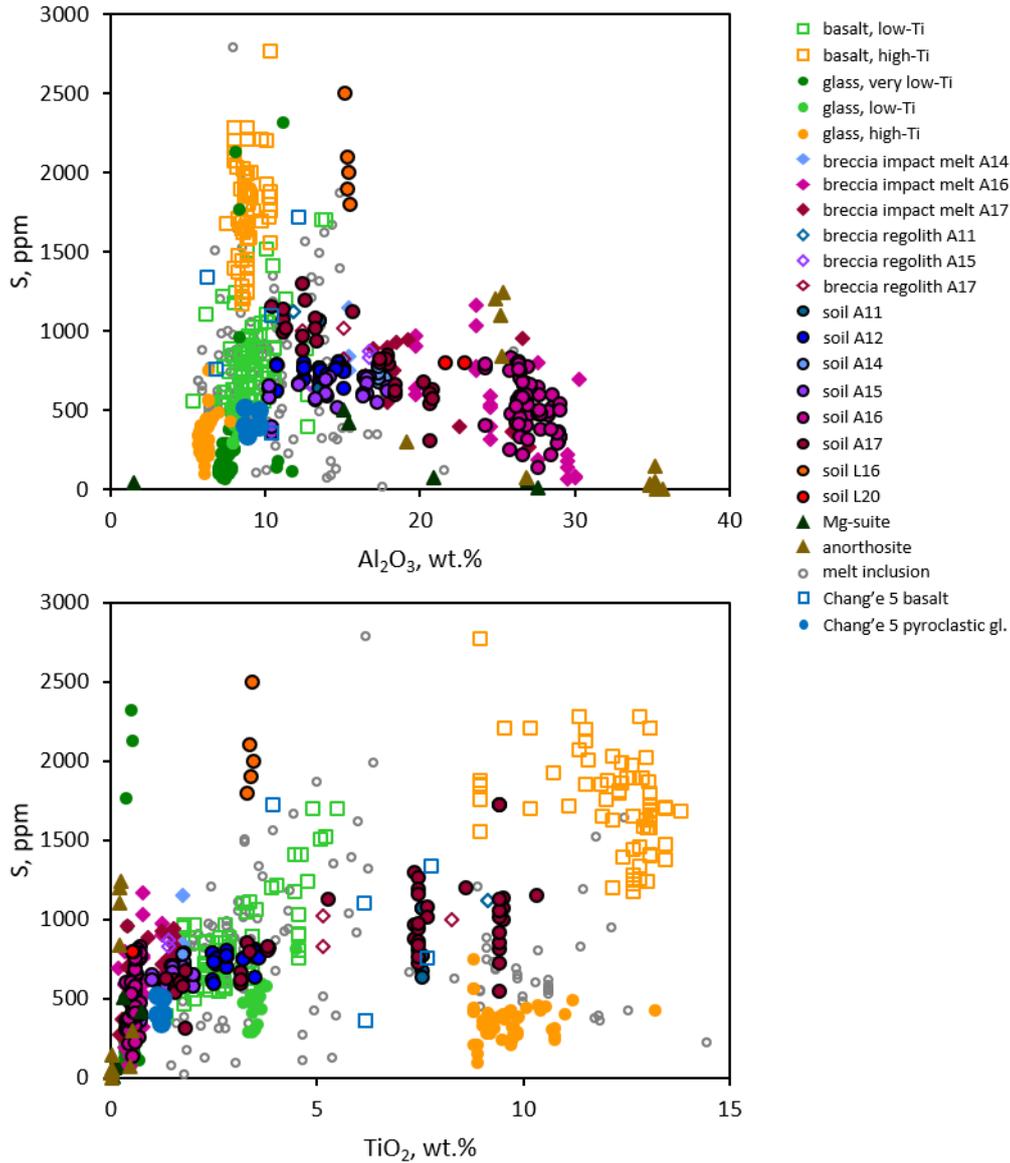

*Figure 3. S concentration (ppm) in lunar rocks as a function of $Al_2O_3$ and $TiO_2$. The data include bulk concentrations of S (and $\delta^{34}S$ values, see Fig. 4) for mare basalts, impact and regolith breccias, Apollo and Luna soils, Mg-suite, and anorthosites, as well as in-situ measurements from pyroclastic glass beads and olivine hosted melt inclusions from mare basalts and pyroclastic glasses. The compilation includes S data determined by combustion, acidification, SIMS, nanoSIMS, and EMPA analysis, and major element concentrations determined by INAA, XRF, SSMS, fused-bead EMPA, AA, and wet chemistry (Kaplan et al., 1970; Kaplan and Petrowski, 1971; Vinogradov, 1971; Thode and Rees, 1971; Rees and Thode, 1972; Gibson and Moore, 1973; Vinogradov, 1973; Chang et al., 1974; Gibson and Moore, 1974; Petrowski et al., 1974; Rees and Thode, 1974; Gibson et al., 1975; Kerridge et al., 1975; J. F. Kerridge et al., 1975; Gibson et al., 1976; Kaplan et al., 1976; Thode and Rees, 1976; Gibson et al., 1977; Des Marais, 1978; Kerridge et al., 1978; Ding et al., 1983; Des Marais, 1983; Norman et al., 1995; Elkins-Tanton et al., 2003b; Bombardieri et al., 2005; Saal et al., 2008; Hauri et al., 2011; Chen et al., 2015; Wing and Farquhar, 2015; Ni et al., 2019; Che et al., 2021; Saal and Hauri, 2021; Liu et al., 2022; Gargano et al., 2022; Gleißner et al., 2022; Renggli, 2023; Wang et al., 2023, 2024).*

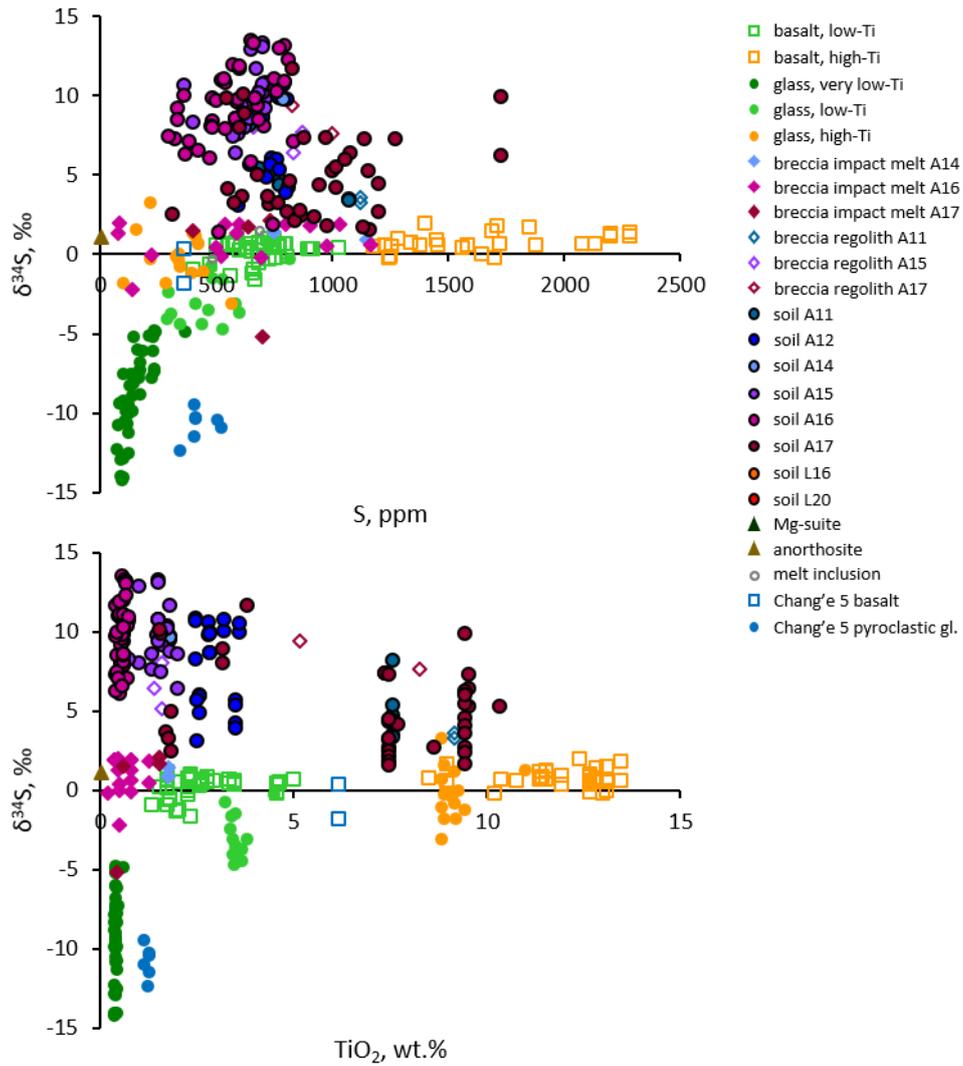

*Figure 4. Stable isotope $\delta^{34}S$ signature in lunar rocks, glasses, and olivine hosted melt inclusions as a function of the S concentration and TiO$_2$. The lunar soils are systematically enriched in heavy S-isotopes, whereas degassed picritic glasses are depleted.*

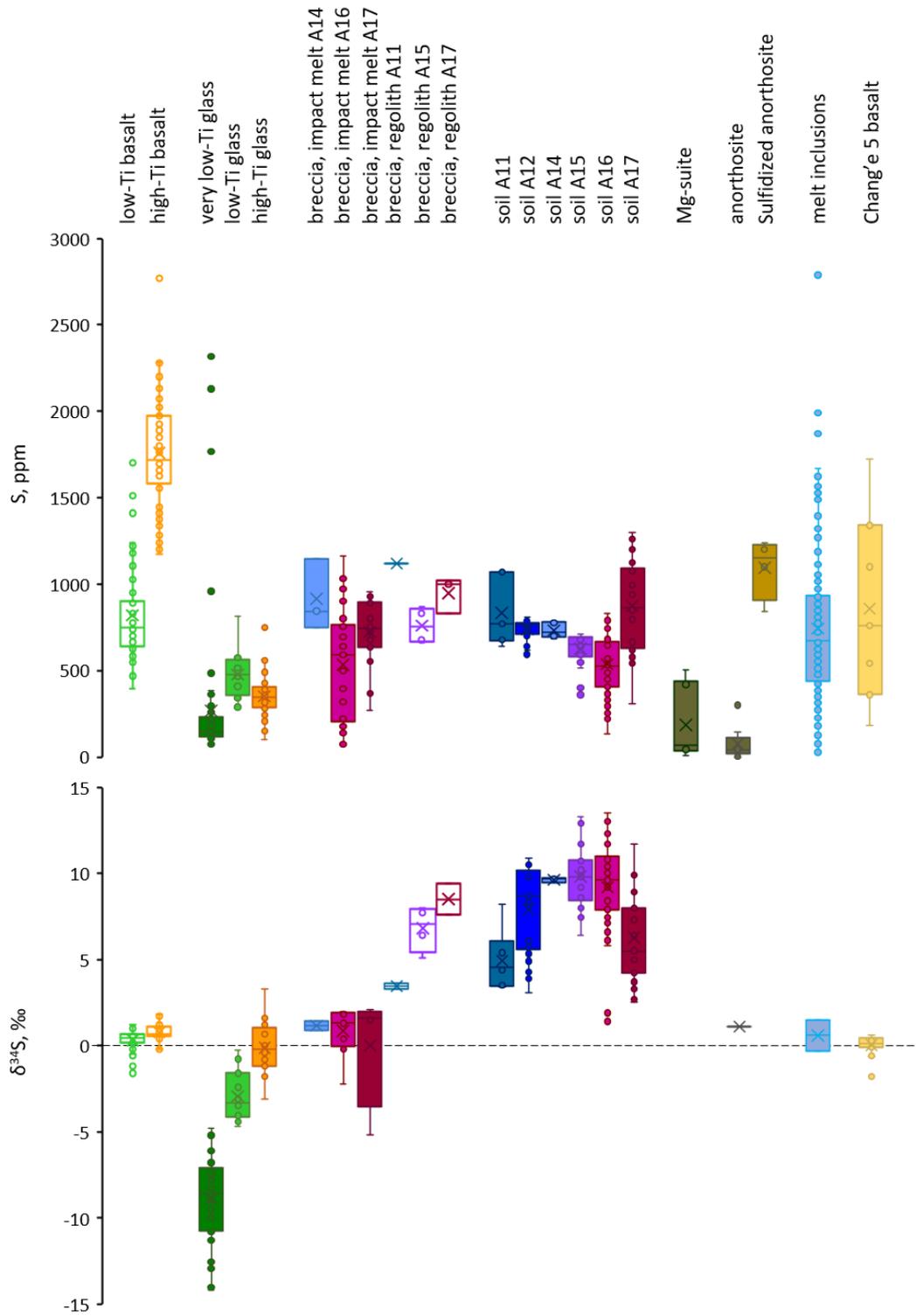

*Figure 5. Box-plot representation of S data from lunar samples. The symbols indicate the first and third quartile (lower and upper limits of the box), the median (horizontal line in box), the average (x symbol), minimum and maximum values (whiskers), and outliers (circles).*

The major element oxide abundances of $Al_2O_3$ and $TiO_2$ allow a clear discrimination between the different lunar sample types (Figure 3). The mare basalts, picritic glasses, and their olivine-hosted melt inclusions are distinguished from breccias and soils based on their lower $Al_2O_3$ concentration. The mare basalts and the picritic glass beads fall in three categories based on their $TiO_2$ content: very low-Ti (<1 wt.% $TiO_2$), low-Ti (1-6 wt.% $TiO_2$), and high-Ti (>6 wt.% $TiO_2$) (Papike et al., 1976; Neal and Taylor, 1992). The high-Ti mare basalts were sampled by Apollo 11 and 17, and the compiled dataset includes data from 5 Apollo 11, and 37 Apollo 17 high-Ti mare basalts (Kaplan et al., 1970; Rees and Thode, 1972, 1974; Gibson and Moore, 1974; Petrowski et al., 1974; Gibson et al., 1975, 1976; Des Marais, 1983; Wing and Farquhar, 2015; Gargano et al., 2022; Gleißner et al., 2022). The high-Ti picritic glasses (orange glasses) were also sampled by Apollo 17 and we include data obtained from the orange soil 74220 and their olivine hosted melt inclusions (Chang et al., 1974; Gibson and Moore, 1974; Thode and Rees, 1976; Ding et al., 1983; Saal et al., 2008; Hauri et al., 2011; Chen et al., 2015; Ni et al., 2019; Saal and Hauri, 2021), and the double drive tube 74001 (Ding et al., 1983). Note that some samples classified under soil A17 (Apollo 17 soils) also include a significant orange glass component. The low-Ti mare basalts were sampled by Apollo 12, 15, and Luna 16. We include data from 27 Apollo 12 mare basalts and from melt inclusions from 7 samples (Kaplan and Petrowski, 1971; Thode and Rees, 1971; Rees and Thode, 1972; Gibson et al., 1977; Bombardieri et al., 2005; Wing and Farquhar, 2015; Ni et al., 2019; Gargano et al., 2022; Gleißner et al., 2022), 11 Apollo 15 mare basalts and from olivine hosted melt inclusions from the samples 15016 and 15647 (Gibson et al., 1975; Kaplan et al., 1976; Des Marais, 1978, 1983; Wing and Farquhar, 2015; Ni et al., 2019; Gargano et al., 2022; Gleißner et al., 2022), and two Luna 16 samples (Vinogradov, 1971, 1973). We also included data from Apollo 14 low-Ti mare basalt 14053 (Gargano et al., 2022). The low-Ti and very low-Ti picritic glasses include the green glass clod samples 15426 and 15427 (Elkins-Tanton et al., 2003a; Saal et al., 2008; Saal and Hauri, 2021). We could not find data on bulk S abundances in very low-Ti mare basalts.

The high-Ti basalts contain on average 1758 ppm S, ranging from 1174 to 2770 ppm. In contrast, the low-Ti basalts contain approximately half of that amount, with 820 ppm S on average, ranging from 396 to 1700 ppm at similar MgO contents (Figure 5). Notably, S concentrations in low-Ti mare basalts correlate with the $TiO_2$ concentration, but S concentrations in high-Ti mare basalts do not (Figure 3). More broadly, the S concentrations in all mare basalts are positively correlated with the incompatible and refractory element Sm, and negatively correlated with Mg# (molar $Mg/(Mg+Fe^{2+})$ ratio) (Gleißner et al., 2022). These correlations are not only observed for S, but also for other siderophile elements, such as Se, Cu, Ag, Co, and Ni, and suggest that the abundances of these elements in the mare basalts are primarily controlled by fractional crystallization of olivine (Steenstra et al., 2018; Gleißner et al., 2022).

The picritic glasses have systematically lower S contents compared to the mare basalts. The glasses cover a broad range of $TiO_2$ concentrations (0.2–17 wt.%) similar to the mare basalts, but have higher Mg# and lower CaO and $Al_2O_3$ contents than most fine-grained, non-cumulate mare basalts, and therefore represent the most primitive magmas (partial melts of the mantle source erupting without significant modification from the initial composition) on the Moon (Shearer and Papike, 1999). The primitive melts originate at a depth of more than 400 km in the lunar mantle (Shearer and Papike, 1993). Even though the glasses experienced volatile loss during the pyroclastic eruption (Heiken et al., 1974; Head and Wilson, 1989; Fogel and Rutherford, 1995; Wilson, 2003; Elkins-Tanton et al., 2003b; Rutherford and Papale, 2009; Wilson and Head, 2017; Rutherford et al., 2017), volatiles are partially retained, as observed by diffusional zonation of S, $H_2O$, Cl, and F in the glass beads (Elkins-Tanton et al., 2003b; Saal et al., 2008; Saal and Hauri, 2021). The average S content retained in the high-Ti orange glasses is 351 ppm (100 to 750 ppm), and the content in the low-Ti and very low-Ti green glasses is 476 ppm (289 to 816 ppm) and 267 ppm (67 to 2317 ppm) respectively (Figure 5).

The available S abundance data in unmodified ferroan anorthosites is very limited. This is primarily due to the low S concentration in these rocks, which was below the detection limits of the methods available during the Apollo era. The ferroan anorthosites largely consist of calcic plagioclase, with very few S-bearing minor phases. Recently, Gleißner et al. (2022) measured the S concentration in 5 Apollo 15 and 16 samples (15415, 60025, 61015, 62255, 65315), with an average of 40 ppm. The S concentration in the ferroan anorthosite 60025 was previously determined at 145 ppm (Des Marais, 1983). In an anorthosite sampled by Luna 20, 300 ppm S were measured (Vinogradov, 1973). The highest S concentration in a ferroan anorthosite clast was measured in the Apollo 16 breccia 67016, with an average of 1095 ppm (Norman, 1981; Norman et al., 1995). The high abundance of troilite in this sample is unusual and likely the result of secondary S addition by a fumarolic gas (see below).

The S concentration in Mg-suite rocks is negatively correlated with the Mg#, which suggests that the S abundance was controlled by fractional crystallization of olivine (Gleißner et al., 2022). The Mg-suite rocks with the highest S abundances and lowest Mg# are the gabbronorite 76255 (419 ppm S, Mg# = 67) and norite 77215 (504 ppm S, Mg# = 69), in contrast to the S-poor high-Mg# dunite 72415 (44 ppm S, Mg# = 87), anorthositic norite 15455 (63 ppm S, Mg# = 81), troctolitic anorthosite 76335 (9 ppm S, Mg# = 88), and norite 78235 (71 ppm S, Mg# = 81) (Gibson and Moore, 1974; Gleißner et al., 2022).

The compositions of the soils and breccias sampled at different Apollo and Luna landing sites mirror those of the locally dominant mare basalts, pyroclastic deposits, or anorthosites. For example, the highest $Al_2O_3$ concentrations of around 30 wt.% are measured in anorthosites, and the soils and breccias from the Apollo 16 and Luna 20 landing sites, located in anorthosite-rich terrain. Similarly, this is also observed in the S abundance in soils and breccia, where the lowest S abundances are measured in Apollo 16 soils (536 ± 170 ppm) and impact melt breccias (534 ± 310 ppm) (Figure 3). The most S-rich soils sampled by Apollo are those from the Apollo 11 (833 ± 174 ppm on average), and Apollo 17 (874 ± 246 ppm) landing sites, and the most S-rich breccias are regolith breccias from Apollo 11 (1120 ppm, 10060,22 Kaplan et al. 1970), and soils (949 ± 86 ppm) as well as impact melt breccias from Apollo 17 (722 ± 195 ppm). This coincides with the prevalence of S-rich high-Ti basalts at the Apollo 11 and Apollo 17 landing sites (Figure 5).

Data on the $\delta^{34}S$ stable isotope composition of lunar samples is more limited compared to data on the S abundance. In Figure 4 we show the $\delta^{34}S$ composition of lunar samples as a function of S and $TiO_2$ content. Broadly speaking, the mare basalts have a positive $\delta^{34}S$, close to zero, the picritic glasses have negative $\delta^{34}S$, the impact melt breccias have slightly positive $\delta^{34}S$, and the soils and regolith breccias have high positive $\delta^{34}S$ compositions.

The $\delta^{34}$S compositions of the mare basalts fall within a narrow range. The high-Ti basalts have an average $\delta^{34}$S composition of 0.77‰ (-0.2 to 2‰), and the low-Ti basalts an average $\delta^{34}$S composition of 0.31‰ (-1.6 to 1.2‰). Overall, the lunar mare basalts are isotopically homogeneous with a mean isotopic composition of $\delta^{34}$S = 0.58 ± 0.05‰, $\Delta^{33}$S = 0.008 ± 0.006‰, and $\Delta^{36}$S = 0.2 ± 0.2‰ (Wing and Farquhar, 2015). No significant $\delta^{34}$S variation occurs as a function of the bulk S concentration in the different mare basalts (Figure 4). The $\delta^{34}$S isotopic compositions of troilites in Chang'e-5 basalts show significantly more variations from 2 to -1.6‰, with a calculated bulk isotopic composition of $\delta^{34}$S = 0.35 ± 0.25‰ (Liu et al., 2022; Wang et al., 2024).

The highest $\delta^{34}$S in lunar samples were measured in soil samples. High positive $\delta^{34}$S values are systematically observed in all soils sampled by the Apollo missions, with the highest values measured in Apollo 15 and 16 soils (13.3‰ and 13.5‰, respectively). The average $\delta^{34}$S compositions of the soils are: Apollo 11 $\delta^{34}$S = 4.93 ± 1.61‰ (n = 6), Apollo 12 $\delta^{34}$S = 7.90 ± 2.58‰ (n = 22), Apollo 14 $\delta^{34}$S = 9.64 ± 0.12‰ (n = 3), Apollo 15 $\delta^{34}$S = 9.82 ± 1.85‰ (n = 21), Apollo 16 $\delta^{34}$S = 9.20 ± 2.61‰ (n = 43), Apollo 17 $\delta^{34}$S = 6.26 ± 2.57‰ (n = 19) (Figure 5). The lack of correlation between the $\delta^{34}$S isotopic composition and surface solar wind exposure ages suggests that the $\delta^{34}$S signatures are not related to solar wind implantation (Chang et al., 1974). There is also no significant difference in S concentrations between fines from permanently shadowed regions, buried soil samples, or surface sites (Gibson and Moore, 1973).

The $\delta^{34}$S of Apollo 14, 16, and 17 melt breccias fall in a similarly narrow range with average compositions close to those of the mare basalts, with 1.17 ± 0.27‰, 0.87 ± 1.14‰, and 0.03 ± 3.02‰ respectively. In contrast, the regolith breccias are significantly more enriched in heavy S, with $\delta^{34}$S compositions of the Apollo 11, 15, and 17 regolith breccias of 3.45 ± 0.15‰, 6.80 ± 1.15‰, and 0.03 ± 3.02‰, respectively. However, note that the available data on $\delta^{34}$S compositions of lunar breccias is very limited and the isotopic compositions are likely more varied. Only one $\delta^{34}$S value is reported for ferroan anorthosites with sample 60025,384: $\delta^{34}$S = 1.1 ± 0.3‰ (Des Marais, 1983). Des Marais (1983) also noted that variation in lunar $\delta^{34}$S values is due to systematic differences in analytical techniques used among laboratories.

**Mare Basalts**

The primary S-bearing mineral in mare basalts is the sulfide troilite (FeS). With less than 1 wt.% the minor elements in troilites from mare basalts are Ti, Co, Ni, and Cr (Papike et al., 1991; McCubbin et al., 2015). Magmatic troilite occurs as a late-stage interstitial phase (Skinner, 1970), or in last-stage melt pockets (mesostasis) (Figure 6). In the mesostasis regions, enriched in incompatible elements, troilite is a minor phase and commonly associated with apatite, silica, and K-rich glass (Potts et al., 2016). Troilite was also identified in mesostasis regions in the youngest lunar mare basalts (~2 Ga) from the Chinese Chang'e-5 sample return mission, with very minor chalcopyrite ($CuFeS_2$), cubanite ($CuFe_2S_3$), and pentlandite ($(Fe,Ni)_9S_8$) (Liu et al., 2022). Minor sulfides including chalcopyrite, cubanite, and pentlandite, as well as mackinawite ($Fe_{1+x}S$), sphalerite (ZnS), and bornite ($Cu_5FeS_4$) had previously been identified in lunar basalts (Levinson and Taylor, 1971; Haskin and Warren, 1991; Papike et al., 1991).

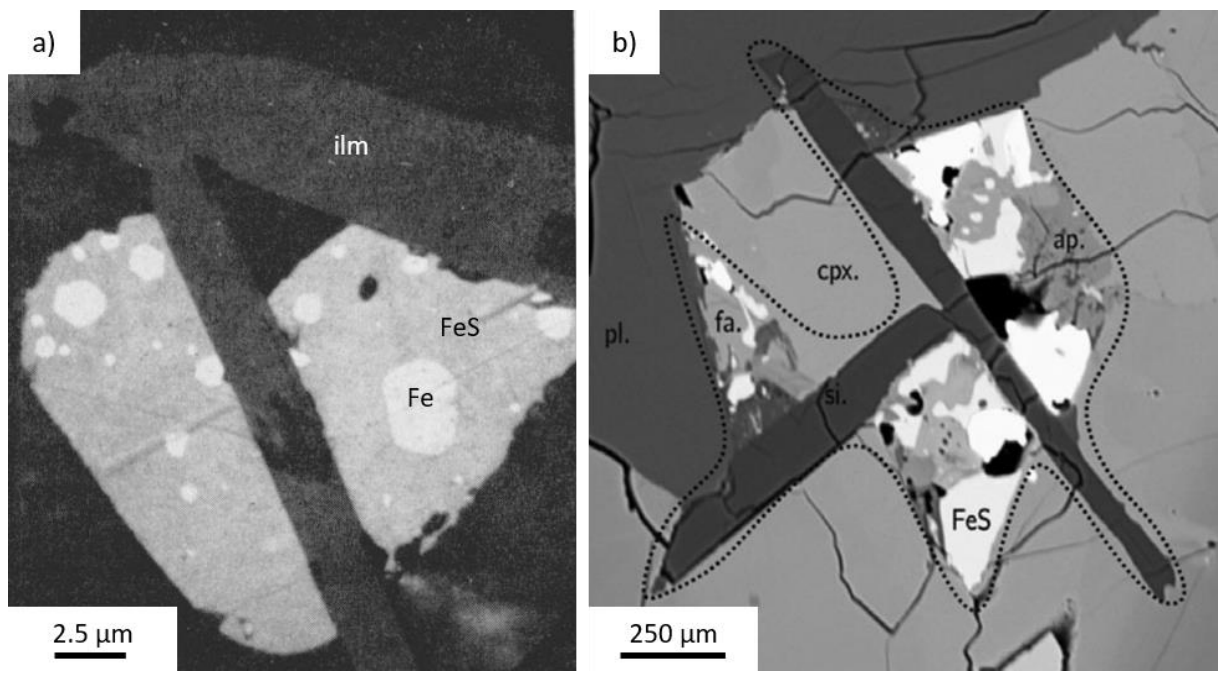

*Figure 6. a) Troilite with Fe metal as an interstitial phase between ilmenite crystals in 10072 (reproduced from Skinner, 1970). b) Troilite with apatite (ap.), fayalite (fa), and silica (si.) in a mesostasis region surrounded by clinopyroxene (cpx.) and plagioclase (pl.) (reproduced from Potts et al., 2016).*

Troilite is commonly associated with metallic Fe, for example shown in Figure 6a in a high-K ilmenite basalt (10072) (Skinner, 1970). Skinner (1970) suggested that troilite always occurs with metallic Fe, as the product of immiscibility of an initially homogeneous melt in the Fe-S system with less than the eutectic S concentration (44 mol% S). On the other hand, the S concentration in mare basalts is negatively correlated with the metallic Fe content of these rocks (Brett, 1976). This was interpreted as the product of S loss either by degassing from a melt ($S^{2-}_{(melt)} + Fe^{2+}_{(melt)} = \frac{1}{2}S_{2(gas)} + Fe^0$) or the decomposition and desulfuration of troilite ($FeS = \frac{1}{2}S_{2(gas)} + Fe^0$) (Gibson and Moore, 1974; Gibson et al., 1975; Brett, 1976).

The accessory mineral apatite is a host of S in lunar basalts, primarily occurring in mesostasis regions (Boyce et al., 2010; 2014). Boyce et al. (2010) reported up to 463 ± 43 ppm S in the high-Al basalt 14053. Recently, more extreme variations in S concentrations (20 ppm to 2800 ppm) and $\delta^{34}$S compositions (-33 ‰ to +36 ‰) were reported for lunar apatites in a preprint (Faircloth et al., 2020). Li et al. (2024) measured S concentrations of up to 3000 ppm in lunar apatites in Chang'e 5 samples, positively correlated with Cl contents. The S contents and isotopic compositions of lunar apatites may be records for degassing processes in the lunar crust (Ustunisik et al., 2011, 2015; Brounce et al., 2014). In contrast to terrestrial apatites, which contain oxidized $S^{6+}$, lunar apatites contain $S^{2-}$ as observed by in-situ S-XANES measurements (Brounce et al., 2019). Possible secondary terrestrial oxidation of S in lunar apatites will be tested by investigating pristine Apollo 17 samples from the NASA Apollo Next Generation Sample Analysis (ANGSA) program (Brounce et al., 2020).

The S abundance in mare basalts correlates with indicators of olivine fractional crystallization, and Gleißner et al. (2022) argued for limited S degassing on this basis. In addition, the isotopic homogeneity of the mare basalts in terms of $\delta^{34}$S is evidence for the lack of substantial S-degassing from mare basalts upon eruption, though degassing without, or small modification of $\delta^{34}$S is possible, depending on the speciation of the gas phase (Wing and Farquhar, 2015), or in the presence of a S-poor Fe-sulfide liquid (Brenan et al., 2019).

As an example of mare basalts, where the degree of degassing remains a topic of discussion, we consider the Apollo 12 low-Ti olivine basalts (e.g. 12002, 12006, 12009, 12012, 12014, 12015, 12018, 12020, 12036, 12040 (Neal et al., 1994)). Bombardieri et al. (2005) analyzed S concentrations

in olivine-hosted melt inclusions from a sub-set of these basalts and determined a primary magmatic composition of 1050 ppm S from melt inclusion concentrations of 824 ± 176 ppm. Based on a comparison with literature data on the bulk concentration of S in these rocks, they determined S-degassing upon eruption of up to 60% (Bombardieri et al., 2005). A comparison with our broader compilation of published S concentrations in whole rock analysis of low-Ti olivine basalts (Figure 5) shows S concentrations of 693 ± 151 ppm (Thode and Rees, 1971; Rees and Thode, 1972; Gibson et al., 1977; Wing and Farquhar, 2015; Gargano et al., 2022; Gleißner et al., 2022). The most recent data exists for sample 12020. Gleißner et al. (2022) measured a S concentration of 657 ± 85 ppm (Mg# = 53.5), which is the same as the S concentration measured in olivine hosted melt inclusions from sample 12020 at 723 ± 159 ppm with an average Mg# = 61.5 (e.g. the olivine hosted melt inclusions are more primitive than the bulk basalt) (Bombardieri et al., 2005). However, a comparison with the inferred primary magmatic composition (1050 ppm, Bombardieri et al. 2005) still suggests a loss of 37 ± 8% of the initial S for basalt 12020, and 34 ± 14% for the entire suite of Apollo 12 low-Ti olivine basalts.

Liu et al. (2022) presented S concentration and $\delta^{34}S$ isotopic compositions from the young Chang'e-5 basalts. The S concentration was calculated from troilite modal abundances, with 360 ± 180 ppm, which is lower compared to S concentrations measured in low- and high-Ti mare basalts returned by the Apollo missions (Figure 5). The Chang'e-5 basalts show evidence of S loss by degassing, estimated at 40 % of the initial S (~600 ± 300 ppm) based on isotopic fractionation for some fragments (Liu et al., 2022). However, based on a larger sample of troilites in basalt clasts with $\delta^{34}S = 0.35 ± 0.25$ ‰, compared to the estimated primary magma composition ($\delta^{34}S = 0.5 ± 0.3$ ‰), S degassing was likely very limited from the bulk Chang'e 5 basalts, and the low S abundance in the basalt is representative of a low S concentration in the mantle source (Wang et al., 2024). Furthermore, the S isotopic composition of the young (2.0 Ga) Chang'e 5 basalts is very similar to much older lunar basalts (Wang et al., 2024).

**Pyroclastic glasses and lunar soils**

The discussion above shows that the degree of S-loss by degassing from erupting mare basalts remains a topic of debate. In contrast, the evidence for substantial degassing of S and other volatile and moderately volatile elements from picritic pyroclastic glasses is unequivocal (Weitz et al., 1999; Saal et al., 2008; Hauri et al., 2015; Wetzel et al., 2015; Saal and Hauri, 2021; Head et al., 2023). Volatile concentration profiles in very-low-Ti Apollo 15 glasses show diffusion trends to the bead surface, which is evidence for degassing of the volatiles H, Cl, F, and S (Figure 7). Numerical modeling of diffusive degassing suggests pre-eruptive water contents of up to 745 ppm, where 98% of the initial water content was lost. Sulfur is less degassed compared to $H_2O$, Cl, and F, at a loss rate of 19% (Saal et al., 2008). Carbon degassed as CO at higher pressures from a depth of more than 4 km, driving the pyroclastic eruptions of the picritic glasses (Head and Wilson, 1989; Weitz et al., 1999; Wetzel et al., 2015). The 74220 orange glasses (Saal et al., 2008; Hauri et al., 2011; Chen et al., 2015; Saal and Hauri, 2021) have a S concentration of $345 \pm 64$ ppm and the olivine hosted melt inclusions a S concentration of $573 \pm 159$ ppm (Hauri et al., 2011; Chen et al., 2015; Ni et al., 2019; Saal and Hauri, 2021). Using these average concentrations, we can calculate that the orange glasses lost $40 \pm 10\%$ of their S concentration by degassing upon eruption. This is similar to the S loss determined for the Apollo 12 low-Ti mare basalts. The primary species in a lunar volcanic gas, with its composition derived from the volatile elements (S, H, C, Cl, F) measured in low-Ti green glass beads (Saal et al., 2008; Wetzel et al., 2015), are $S_2$, $H_2$, and CO, followed by HF, $H_2S$, COS, $CS_2$, and HCl (Renggli et al., 2017; Renggli et al., 2023a). The S concentration data from very low-Ti green glasses shows notable outliers at high concentrations with up to 2300 ppm (Figure 5). These data are from vesicular rims on glass beads in the sample 15426,72 (Elkins-Tanton et al., 2003b). Elkins-Tanton et al. (2003b) argued that the high S concentration in the rims suggests that they did not degas and the melts originated at a pressure of 2.2 GPa, where they may have been S saturated. However, the unique nature of the S-rich rims demands some caution, without observations in other similar samples. Varnam et al. (2024) reviewed models for the compositions of lunar volcanic gases for pyroclastic glasses and mare basalts, based on data from the literature (Tartèse et al., 2013; Needham and Kring, 2017; Newcombe et al., 2017; Rutherford et al., 2017; Kring et al., 2021; Renggli et al., 2023). This compilation showed that the most important likely S-bearing gas species in lunar volcanic gas are $H_2S$ and $S_2$. Varnam et al. (2024) further concluded that likely $H_2$ was the most abundant molecular species in lunar volcanic gas. In terms of elemental abundances H is followed by O, S, and C (Varnam et al., 2024).

There is a very limited number of $\delta^{34}S$ measurements in olivine hosted melt inclusions. Presently, there are only three reported $\delta^{34}S$ measurements in olivine-hosted melt inclusions from the high-Ti picritic glasses in sample 74220, ranging from +1.6 to −0.3‰, compared to the $\delta^{34}S$ values varying from +1.3 to −1.8‰ in the high-Ti glasses (Hauri et al., 2011; Ni et al., 2019; Saal and Hauri, 2021).

Apollo 15 green glass beads are vitrophyric and only contain occasional olivine crystals. Based on the olivine morphologies and grain sizes, cooling rates were estimated at 1 °C/s (Arndt et al., 1984). If cooling occurred during free flight, in the absence of a volcanic gas surrounding the beads, a much higher cooling rate of 1500–4200 °C/s would be expected. The lower cooling rate is clear evidence for the presence of a hot gaseous medium (Arndt et al., 1984; Renggli et al., 2017). The cooling rate of Apollo 17 orange and black glass beads is estimated at 100 °C/s, which is two orders of magnitude faster than that of the green glasses, but still requires the presence of a hot gas (Arndt and von Engelhardt, 1987). Direct measurements of cooling rates of lunar orange glass beads (74220,867) using heat capacity-temperature paths show a significantly slower rate at 1.7 °C/s (Hui et al., 2018), closer to that estimated by Arndt et al. (1984) for the Apollo 15 green glasses.

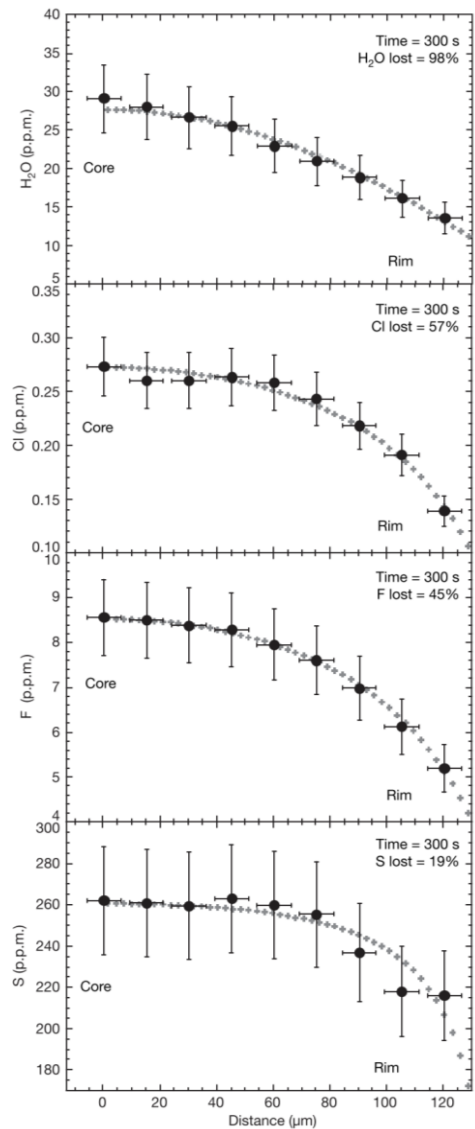

*Figure 7. Concentration profiles from core to rim in a picritic glass bead (Green #5) (Saal et al., 2008). The black circles are concentration data measured with a Cameca IMS 6f and a NanoSIMS 50L. The grey crosses show a diffusion model fit describing the diffusive volatile loss at the bead surface (reproduced from Saal et al., 2008).*

An important property of the pyroclastic glass beads is that volatiles are concentrated in coatings on the surfaces of the glass beads. This is exemplified by a compilation of grain-size dependent S data of high-Ti orange glasses from the double drive tube 74001/2 and orange soil 74220 (Rees and Thode, 1974; Thode and Rees, 1976; Ding et al., 1983) shown in Figure 8 (Saal and Hauri, 2021). A surface-correlated increase in volatile element abundances was first identified for Pb on orange glass beads (Tatsumoto et al., 1973). The 74220 orange glass soil has a bulk S concentration of 481 ± 154 ppm, which is intermediate between the in-situ glass concentration, and the concentration measured in melt inclusions. Bulk-sample δ$^{34}$S values of the lunar high-Ti volcanic glasses (74220 and drive tube 74001/2) extend to significantly lighter values, with δ$^{34}$S from +0.69 to -2.6‰ (Rees and Thode, 1972; Thode and Rees, 1976; Ding et al., 1983). These studies found that sulfur concentration increases and δ$^{34}$S values decrease with decreasing grain size fraction of the sample analyzed (Figure 8). The unusual light δ$^{34}$S values at high S content of the volcanic glasses indicate the presence of a surface component produced by the condensation of a volcanic gas cloud onto the surface of the glass beads (Rees and Thode, 1972; Thode and Rees, 1976; Ding et al., 1983). Furthermore, Ding et al. (1983) found that S concentration and δ$^{34}$S decrease with increasing depth within the drive tube, suggesting more degassed volcanic glasses with depth. Finally, an extensive study of 3000 glass beads from the Chang'e 5 samples revealed 3 beads with volcanic pyroclastic origins. These glass beads have S concentrations of 401 – 522 ppm and δ$^{34}$S compositions of -9.5 to -12.4 (Wang et al., 2024). Notably, these pyroclastic glass beads have ages of only 123 ± 15 million years, suggesting very recent explosive volcanism on the Moon (Wang et al., 2024).

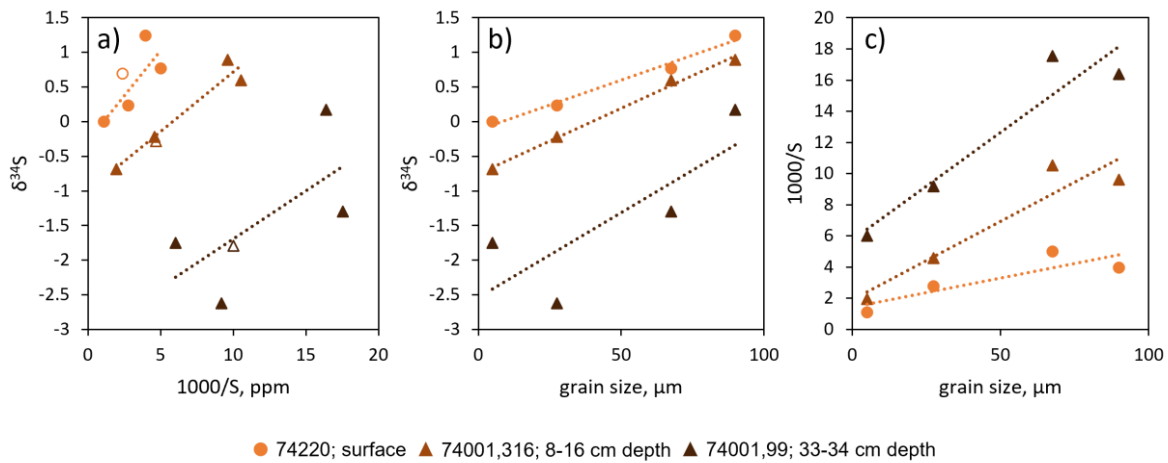

*Figure 8: δ$^{34}$S (‰) bulk-sample analyses of the high-Ti volcanic glasses from sample 74220 and drive tube 74001. a) δ$^{34}$S (‰) vs 1000/S (ppm$^{-1}$), b) and c) δ$^{34}$S (‰) and 1000/S (ppm$^{-1}$) vs. grain size fraction (μm). Circles are from sample 74220, triangles (8-15 cm depth) and diamond (averages of 33-34 and 63-69 cm depth) from drive tube 74002/1. The open symbols in a) represent the calculated bulk compositions. Dotted lines are linear regressions of the data. Y intercepts of the regression lines in b and c represent the δ$^{34}$S value and S content at infinite ratio of surface area to volume; i.e., the composition of the sulfur-rich coatings on the surfaces of the glass beads. The δ$^{34}$S and S content of the coating decreases with the collection depth in the drive tube, consistent with progressively more degassed samples with depth. Samples contaminated with mature regolith have not been considered. Data from Ding et al., 1983 (modified from Saal and Hauri, 2021).*

In addition to S, coatings on pyroclastic glass beads contain other chalcogens including Se and Te, halides F, Cl, Br, and I, alkalis Na and K, and volatile to moderately volatile metals Cu, Zn, Ga, Ge, Ag, Cd, In, Hg, Tl, Pb, and Bi, as well as P and Au (McKay et al., 1973; Tatsumoto et al., 1973; Grant et al., 1974; Heiken et al., 1974; Chou et al., 1975; Jovanovic and Reed, 1975; Meyer et al., 1975; Butler and Meyer, 1976; Goldberg et al., 1976; Wasson et al., 1976; Butler, 1978; Cirlin et al., 1978; Clanton et al., 1978; Cirlin and Housley, 1979; Hauri et al., 2015; Ma and Liu, 2019; Parman et al., 2020; Liu and Ma, 2022; McCubbin et al., 2023). At average coating thicknesses of less than 0.1 μm and individual coating grain sizes of up to 1 μm on the glass beads, few distinct phases have directly been identified (Figure 9). These include ZnS (Butler and Meyer, 1976; Wasson et al., 1976; Butler, 1978) and possibly FeS, CuS, and Ni-sulfide based on element ratios (Butler and Meyer, 1976), and NaCl (Clanton et al., 1978; McKay and Wentworth, 1992; Parman et al., 2020). Secondary alteration and degradation of the phases in the coatings since sample collection in the early 1970s makes an unambiguous identification challenging, though recent investigations suggest the deposition of alkali sulfides (McKay and Wentworth, 1992; Ma and Liu, 2019; Liu and Ma, 2022). Thermodynamic calculations suggest that the metals were transported as atomic (Zn, Pb, Fe, Cu, Ni), sulfide (Pb, Fe, Ni), and chloride (Ga, Cu) gas species prior to deposition as primarily sulfide phases (ZnS, PbS, FeS, CuS, NiS, and GaS) (Renggli et al., 2017). The relationship between element abundances in the coatings compared to the glass composition of the 74220 orange glass beads suggests degassing of 63% of S, compared with 42% Na, 97% Zn, 36% Ga, 93% Pb, and 100% Ge (Hauri et al., 2015).

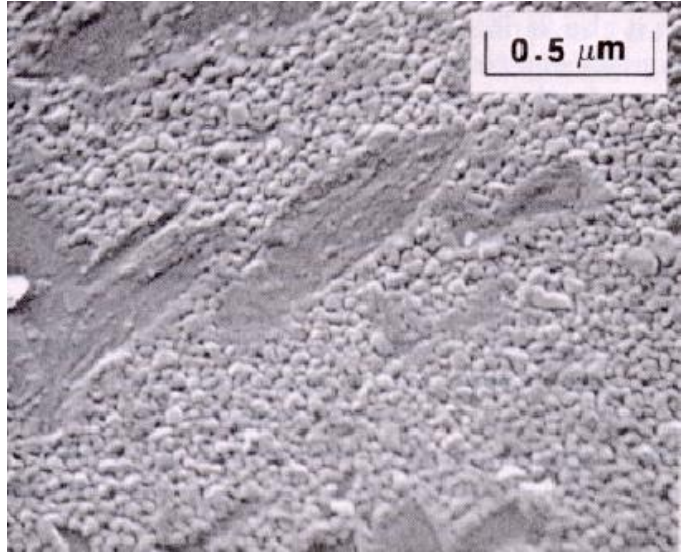

*Figure 9. Secondary electron microscopy image of partially abraded coatings on a high-Ti picritic glass bead from sample 74001,98. The coatings include sulfides and chlorides, and are enriched in volatile and moderately volatile elements (S, Se, Te, F, Cl, Br, I, Na, K, Cu, Zn, Ga, Ge, Ag, Cd, In, Hg, Tl, Pb, Bi) (reproduced from Clanton et al., 1978).*

Saal and Hauri (2021) recently reported *in-situ* measurements of sulfur isotope ratios dissolved in the very low- and low-Ti glasses from Apollo 15 (15426 and 15427) and the high-Ti glasses from Apollo 17 (74220). The new data reveal large variations in $^{34}S/^{32}S$ ratios, which positively correlate with S and Ti contents within and between the distinct compositional groups of volcanic glasses analyzed. They conclude that each compositional group of volcanic glasses have been affected by magmatic Rayleigh degassing of $H_2S$, COS, and $S_2$ gas species suggested by thermodynamic calculations for lunar volcanic gas speciation (Renggli et al., 2017), producing positive correlations between S content and $\delta^{34}S$ values. Furthermore, Saal and Hauri (2021) showed that kinetic isotope fractionation during condensation of S species from the volcanic gas on the surfaces of the glass beads is the most likely explanation for the light $\delta^{34}S$ values of the surface-correlated S content of the high-Ti volcanic glasses (Thode and Rees, 1976; Ding et al., 1983) (Figure 8). Saal and Hauri (2021) propose that the decreasing initial $\delta^{34}S$ and S values with decreasing $TiO_2$ content for the three compositional groups of volcanic glasses analyzed, is consistent with the generation of the lunar basalts during melting of a heterogeneous lunar mantle, generated during core segregation, crystallization of LMO, and overturn and mixing of the cumulate pile. Whether the Earth's and Moon's interiors share a common $^{34}S/^{32}S$ ratio remains a matter of debate.

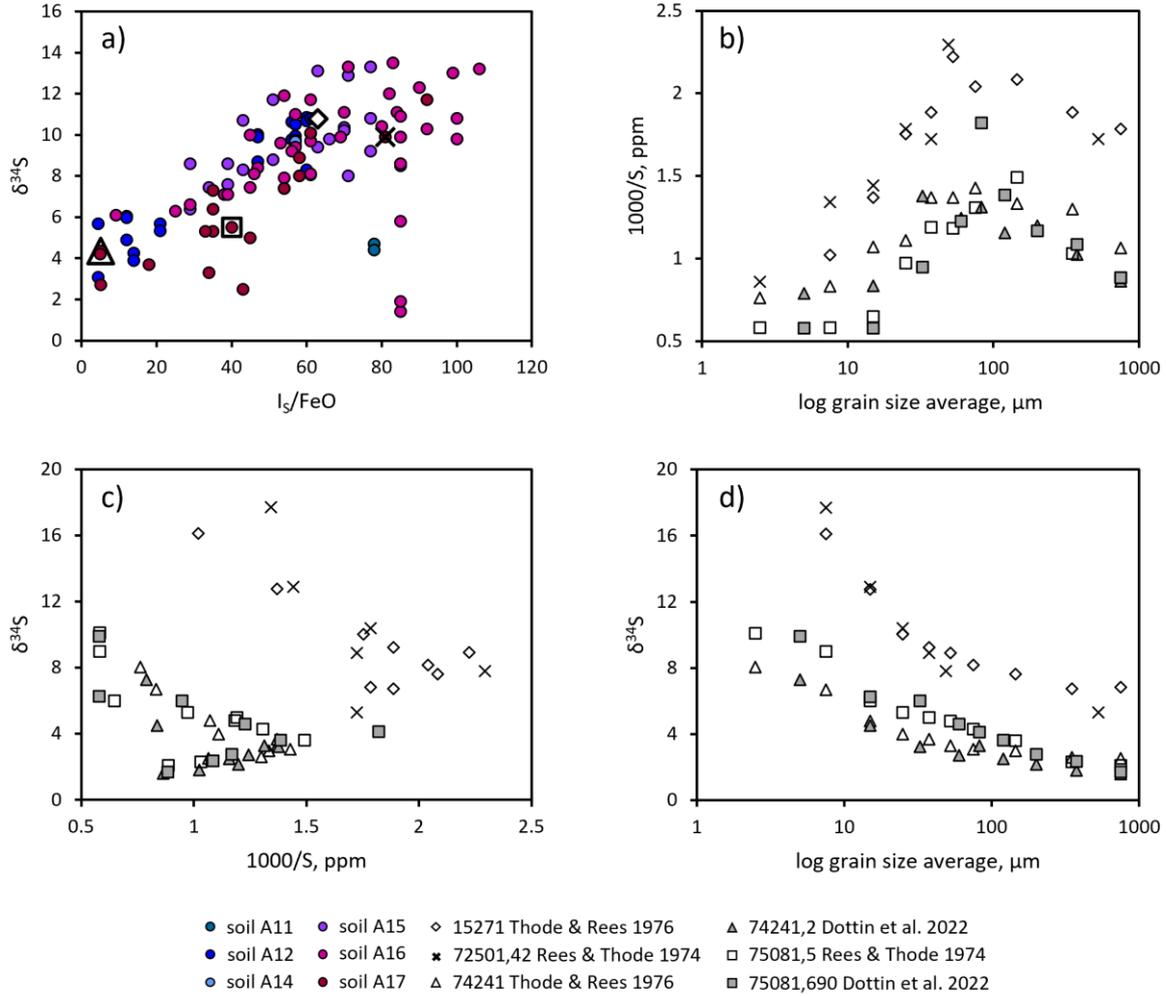

*Figure 10: Correlation of $\delta^{34}S$ (‰) and S (ppm) with grain size (µm) and index of maturity ($I_S/FeO$) in lunar regolith. a) bulk-sample $\delta^{34}S$ (‰) vs. index of maturity ($I_S/FeO$, concentration ratio of nanophase metallic iron to total iron) for a full compilation of lunar regolith data. Regolith samples 72501 (x-symbols), 15271 (diamonds), 75081 (squares), and 74241 (triangles) are highlighted. c) $\delta^{34}S$ (‰) versus 1000/S (ppm$^{-1}$) for different grain size fractions of the highlighted samples. b) and d)1000/S (ppm$^{-1}$) and $\delta^{34}S$ (‰) vs. grain size fraction (µm) analyzed of regolith samples 72501, 15271, 75081, and 74241 showing a variable range of maturity ($I_S/FeO$). Data for panels b-d from (Thode and Rees, 1976; Chang et al, 1974; Dottin et al., 2022). Data for panel a from (Thode and Rees, 1971; 1976; Kaplan et al., 1979;1976; Kaplan and Petrowski 1971; Rees and Thode, 1974; Chang et al, 1974;Pietrowski et al 1974; Kerridge et al., 1975a; ; Kerridge et al., 1975b ; Kerridge et al., 1978) (modified and extended from Saal and Hauri, 2021).*

In contrast to the volcanic glasses, the $\delta^{34}$S value of regolith soil samples increases with S abundance, index of maturity $I_S$/FeO (concentration ratio of nanophase metallic iron to total iron), exposure ages, and decreasing grain size fraction (Figure 10). Several hypotheses have been proposed to explain this systematics: solar-wind sputtering, proton stripping, micrometeorite impact vaporization, and later condensation, or a combination of these processes (Clayton et al., 1974; J. F. Kerridge et al., 1975; Pieters and Noble, 2016). Release of S in a small impact was observed in the Lunar Crater Observation and Sensing Satellite (LCROSS) experiment, in which the upper-stage of the launch vehicle was crashed into the Moon forming a 25-30m diameter crater and a ballistic ejecta plume (Schultz et al., 2010). After $H_2O$ (5.1 ± 1.4 x $10^{19}$ molecules/$m^2$), $H_2S$ was the second most abundant molecule detected in the ejecta plume at 8.5 ± 0.9 x $10^{18}$ molecules/$m^2$, followed by $NH_3$ (3.1 ± 1.5 x $10^{18}$ molecules/$m^2$) and $SO_2$ (1.6 ± 0.4 x $10^{18}$ molecules/$m^2$) (Colaprete et al., 2010). Importantly, the S-bearing species were more abundant relative to $H_2O$ compared to the ratios found in comets, suggesting a S origin from the regolith (Colaprete et al., 2010; Hurley et al., 2023). Apollo samples have become an important reference to constrain these space weathering processes. The lunar soils have recorded space weathering processes, for example by an accumulation of optically active opaque particles, which are nanophase metallic Fe particles in the weathered rims of soil grains (Pieters and Noble, 2016). The effect of space weathering at the lunar surface is also recorded in the S isotopic composition of soil samples. Mass dependent S isotope fractionation occurs by S escape upon micrometeorite impact and increases with soil maturity (Figure 10a), whereas mass independent S isotope fractionation ($\Delta^{33}$S) occurs via interaction of S in a tenuous lunar atmosphere with UV photons by photodissociation and photo-oxidation (Dottin et al., 2022). This mass independent S isotope fractionation has been observed in Apollo 17 soil samples (e.g., 75081,690) with a decrease of $\Delta^{33}$S with decreasing grain size fractions (Dottin et al., 2022). Impact glass beads returned by the Chinese Chang'e 5 mission show a large range of S concentrations and isotopic compositions with 4 – 2345 ppm S and 0 – 58 $\delta^{34}$S, where $\delta^{34}$S increases with decreasing S concentrations (Wang et al., 2024).

On the 23rd of August 2023 India became the fourth nation to successfully land on the Moon, for the first time with a probe near the south pole. The lander Chandrayaan-3 included a rover, named Pragyan ("wisdom" in Sanskrit), with an alpha particle X-ray spectrometer (APXS) and a laser induced breakdown spectroscope (LIBS) as scientific payloads to measure elemental compositions of the lunar surface near the landing site. The mission included the first in-situ measurement of the surface composition at the lunar south pole, with direct detection of S by the LIBS instrument (Kanu et al., 2024).

**Fumarolic alteration**

The S concentration in pristine anorthosites is very low, as discussed above (Figure 5, Gleißner et al. 2022). However, anorthosites, anorthositic breccias, and Mg-suite breccias show abundant sulfide replacement textures (Ramdohr, 1972; Kerridge et al., 1975; Norman, 1981; Haskin and Warren, 1991; Colson, 1992; Norman et al., 1995; Shearer et al., 2012, 2015; Elardo et al., 2012; Bell et al., 2015). Norman (1981) first identified troilite-enstatite intergrowths replacing olivine in plutonic clasts in the suevitic lunar breccia 67016. Importantly, this sample first provided direct evidence for a volatile-rich phase in the lunar crust. Norman et al. (1995) further investigated the sulfidation textures in ferroan noritic anorthosite clasts in 67016 and calculated a bulk S concentration in these metasomatized anorthosites of 1095 ± 156 ppm, which is a more than 25 times higher S content compared to pristine anorthosites (Figure 5).

Figure 11 shows a backscattered electron image of an olivine modified to troilite and orthopyroxene in a Mg-suite clast in the lunar breccia 67915 (Bell et al., 2015). The fine-grained alteration textures also contain oxides, such as Mg-bearing ilmenite, Cr-spinel, and Fe metal (Bell et al., 2015). The primary sulfide-forming reaction of the Fe-component in the olivine likely follows the reaction (Colson, 1992; Shearer et al., 2012, 2015):

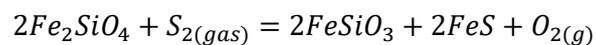

$$2Fe_2SiO_4 + S_{2(gas)} = 2FeSiO_3 + 2FeS + O_{2(g)}$$

In addition to replacement textures, troilite also occurs in veins along grain boundaries and in fractures, often associated with enrichments in moderately volatile elements such as Cu, Zn, Se, Co, and Sb, suggesting the transport of these metals in the gas phase (Krähenbühl et al., 1973; Norman et al., 1995; Shearer et al., 2012). Isotopically, the troilite replacement veins are light, with $\delta^{34}$S values in the range of -1 to -3.3 ‰ (Shearer et al., 2012).

A particularly volatile enriched sample is the well-studied Apollo 16 "Rusty Rock" 66095 (Taylor et al., 1973, 1974; El Goresy et al., 1973; Sharp et al., 2010; Shearer et al., 2014; Day et al., 2017, 2019; Renggli and Klemme, 2021). This sample contains troilite as the primary sulfide, but also sphalerite (ZnS). In addition, the "Rusty Rock" is enriched in Zn, Pb, Cd, Bi, Ge, Sb, Tl, Br, I, and Cl-, and OH-alterations, including the phases akaganéite (FeO(OH,Cl)), and schreibersite ((Fe,Ni)$_3$P) (Krähenbühl et al., 1973; El Goresy et al., 1973; Taylor et al., 1974; Cirlin and Housley, 1980; Shearer et al., 2014). Notably, the alteration and volatile element enrichment in 66095 is not unique among Apollo 16 samples, as rocks from many sampling locations contain alterations (60016, 60235, 60255, 60525, 60625, 61135, 62241, 63585, 64455, 64567, 65095, 65235, 65326, 65359, 65759, 65766, 65779, 66035, 66036, 66055, 66084, 66095, 67455, 68501, 68505, 68841, 69935, 69941, 69961 (Taylor et al., 1973; Jean et al., 2016)). The existence of these alterations within the samples, and sometimes below impact melt coatings, suggests that they are of lunar origin and not due to terrestrial alteration during sample storage (Taylor et al., 1973; Shearer et al., 2014). Shearer et al. (2014) concluded that the gas phase that caused the volatile element enrichments in the altered rocks, did not originate from extra-lunar additions. The fumarolic gas instead likely originates from a degassing ejecta blanket (McKay et al., 1972; Haskin and Warren, 1991; Shearer et al., 2014; Day et al., 2019), or possibly from a degassing shallow intrusion of basaltic melt (Shearer et al., 2014; Day et al., 2017). The fumarolic formation temperature of the "Rusty Rock" alteration assemblage of FeCl$_2$ and (Zn,Fe)S was experimentally constrained to 580 ± 50 °C (Renggli and Klemme, 2021).

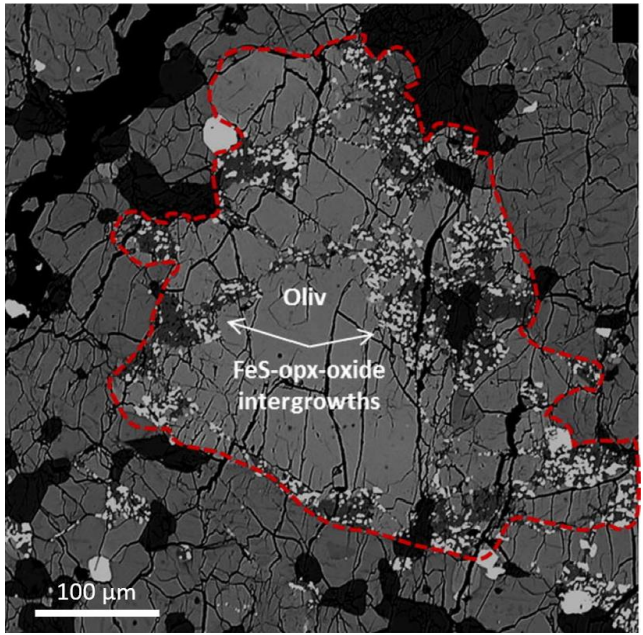

*Figure 11. Backscattered electron image of an olivine phenocryst altered by gas-solid reaction to troilite-orthopyroxene-oxide intergrowths (Bell et al., 2015). Note the preferred reaction pathways along grain boundaries and within cracks of grains (reproduced from Bell et al., 2015).*

### S in the lunar mantle and core

First, we discuss estimates for the S abundance in the bulk silicate Moon (BSM), which makes up more than 98% of the lunar mass, or 96% of the lunar volume (Weber, 2014; Lognonné and Johnson, 2015). The crust, as part of the BSM, makes up 9% of the lunar volume, leaving the lunar mantle reservoir as a large potential S reservoir. Estimates of volatile contents of the bulk lunar mantle are largely based on picritic glasses that are significantly more volatile-rich than crystalline mare basalts previously studied (Hauri et al., 2015; Chen et al., 2015; Ni et al., 2019). To extrapolate the S abundance measured in picritic glasses to lunar mantle abundances, ratios can be determined to non-volatile trace elements whose behavior is well constrained, such as Dy (Saal et al., 2002). For very-low-Ti Apollo 15 green glasses the pre-degassing S/Dy = 129, and for olivine hosted melt inclusions in high-Ti orange glasses 74220 S/Dy = 94.5, compared to S/Dy = 225 ± 50 in depleted MORB (Saal et al., 2002, 2008; Hauri et al., 2015). Assuming that the relationship between S/Dy in depleted MORB to the bulk silicate Earth (BSE) with 250 ppm S (McDonough and Sun, 1995) can be extended to the Moon for an upper BSM estimate, we obtain 116 ± 26 ppm S based on the very-low-Ti Apoll 15 green glasses, and 85 ± 19 ppm based on 74220 olivine hosted melt inclusions. Most calculations of S contents of the bulk lunar mantle based on picritic glasses and melt inclusions (75 ppm, Bombardieri et al., (2005); 78.9 ppm, Hauri et al., (2015); 70 ppm, Chen et al., (2015); 73 ppm, Ni et al., (2019)) fall in the same range as those from bulk mare basalts (72 ppm based on 12009 olivine basalt; 103 ppm based on 15555 olivine-normative basalt, Gleißner et al., (2022). Liu et al. (2022) estimated a lower S abundance in the mantle source of the young Chang'e-5 basalts, at only 1-10 ppm (see also Wang et al., 2024), by accounting for fractional crystallization of a low-degree partial melt. Notably, the Chang'e-5 mantle source is also water-poor (Hu et al., 2021). In contrast, calculations based on volatiles in Chang'e-5 apatites suggest S concentrations in the mantle source in the range of 38 – 125 ppm (Li et al., 2024). See the chapter by Kiseeva & Fonseca (2024, this volume) for a review on S in the Earth's mantle for comparison.

A "hidden" sulfide reservoir, a putative FeS layer at the base of the lunar mantle, has been proposed to have formed by segregation during the late stages of lunar magma ocean crystallization (Morbidelli et al., 2018). Such a basal FeS layer would imply much higher bulk lunar mantle S contents. A "hidden" sulfide reservoir in the deep lunar mantle would also have profound implications for the potential storage of sulfide-loving ("chalcophile") elements, which include Co, Ni, Cu, Se, Te, Pb and Bi (Kiseeva and Wood, 2013; Steenstra et al., 2020a). A hidden sulfide reservoir could only exist if more primitive lunar melts would have been sulfide-saturated at some stage of their evolution. This would have resulted in sulfide segregation during magma ocean solidification, or during subsequent magma petrogenesis from magma ocean cumulates. A number of experimental and Apollo sample studies have been conducted to assess if lunar low- and high-Ti basalts, and volcanic glasses experienced sulfide saturation (Gibson et al., 1976, 1977; Brett, 1976; Danckwerth et al., 1979; Bombardieri et al., 2005; Ding et al., 2018; Steenstra et al., 2018; Day, 2018; Brenan et al., 2019; Gleißner et al., 2022). The latter workers generally concluded from sample analyses, chalcophile element abundances and experimentally measured S solubilities in such melts, that the source regions of the low- and high-Ti basalts and volcanic glasses, and the melts themselves, did not experience sulfide saturation. Hence, the low concentrations of chalcophile elements in the mare basalts are likely a primary feature of the lunar interior (Bombardieri et al., 2005; Steenstra et al., 2018). Brenan et al.

(2019) suggested from highly siderophile element systematics and new S solubility measurements at low oxygen fugacities, that both high- and low-Ti magmas may have been saturated by a S-poor sulfide melt. However, due to the fact that the abundances of highly chalcophile elements (e.g. Ni and Cu) in low-Ti basalts can be explained solely by olivine fractionation (Steenstra et al., 2018; Gleißner et al., 2022), the inferred S-poor sulfide melt must have been rapidly exhausted, providing only a very minor sink for S and most other chalcophile elements.

Sulfide saturation could also have occurred earlier in lunar history, namely during the solidification of the lunar magma ocean (LMO) following formation of the lunar core (Morbidelli et al., 2018). The incompatible behavior of S in mineral-melt systems, in conjunction with lower S solubilities in silicate melts with decreasing temperature, could have resulted in sufficient enrichment of S in the residual magma ocean liquid for sulfide saturation to have occurred at some stage. Steenstra et al. (2020) tested this hypothesis in detail, but found that, even in the absence of S-loss due to degassing, the LMO melts could only plausibly have been sulfide-saturated at the final few % of LMO crystallization. At this stage, it would be very difficult to efficiently recycle such sulfide liquid homogeneously back into the lunar mantle.

Sulfur and C are expected to be the most abundant light elements in the lunar core, given their slightly to moderately siderophile behavior at the pressure-temperature-redox conditions expected for lunar core formation (Steenstra et al., 2016, 2017b; Righter et al., 2017; Steenstra et al., 2020a; Komabayashi & Thompson 2024, this volume). Current estimates of the S content of the lunar core are based on geophysical data, i.e. Apollo-era seismograms (Weber et al., 2011; Garcia et al., 2019) as well as Lunar Laser Ranging data (Smith et al., 2010), the existence of an ancient lunar core dynamo (Cisowski et al., 1983; Shea et al., 2012; Laneuville et al., 2014), and geochemical constraints, such as the lunar mantle depletion of S and that of other iron-loving element depletions (Rai and van Westrenen, 2014; Steenstra et al., 2016, 2017b, a).

Geophysical observations point to the existence of a partially molten outer lunar core (Weber et al., 2011). To sufficiently reduce the liquidus temperature of the lunar core required for sustaining a present liquid state, high S contents have been proposed for the lunar liquid outer core with estimates between 6 to 11 wt.% S (Antonangeli et al., 2015). Numerical models simulating the ancient lunar core dynamo yield similar high S contents of the lunar core (Laneuville et al., 2014). Overall, bulk lunar core estimates of S derived by these approaches span from 3 to 12 wt.% S (Weber et al., 2011; Zhang et al., 2013; Laneuville et al., 2013, 2014; Jing et al., 2014; Antonangeli et al., 2015).

Geochemical evidence generally points to much lower S contents of the lunar core. Rai and van Westrenen (2014) proposed that at least 6 wt.% S would be required for explaining the lunar mantle depletion of V, whereas Steenstra et al. (2016) suggest that siderophile element depletions can also be reconciled with a S-poor lunar core at high temperatures (Figure 12). The iron-loving element depletion pattern is generally consistent with equilibrium between the lunar metal core and the silicate mantle (e.g. Rai and van Westrenen 2014, and references therein), whereas the bulk Moon volatile composition is either similar to the primitive bulk silicate Earth (e.g., in terms of S, Wing and Farquhar, 2015) or much more depleted, i.e. due to volatile loss during formation of the Moon (Kato and Moynier, 2017). If one considers both of these observations in light of plausible S contents of the lunar core, the lunar core must be S-poor. A BSE-like bulk Moon composition yields only <0.14 wt.% S in the bulk lunar core (Steenstra et al., 2017a), and any volatile loss of S during the Moon-forming event would only further decrease the bulk S content of the lunar core. Indeed, consideration of primitive BSE S contents for the bulk Moon, and subsequent depletion of S in the Moon due to core formation, yields lunar mantle S concentrations that are very close or within error of reported estimates for the lunar mantle (70-80 ppm) (Hauri et al., 2015; Chen et al., 2015). For a discussion of S in the lunar core in comparison to the cores of Earth and Mars see Komabayashi & Thompson (2024, this volume).

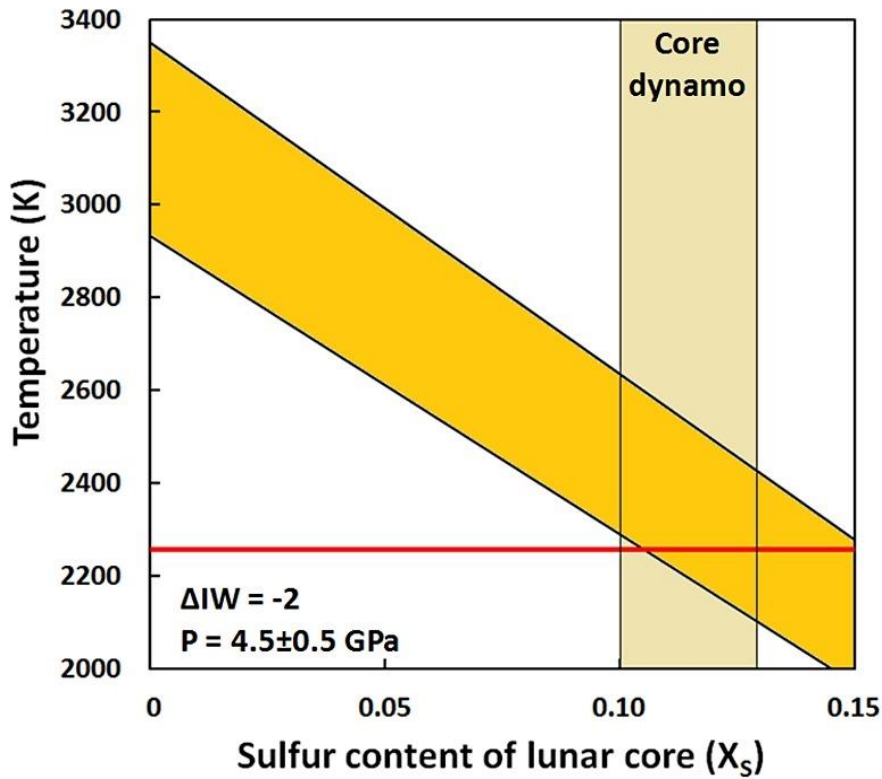

*Figure 12. The modeled lunar core sulfur content and temperature (indicated by the filled orange range within solid lines) at which non-volatile iron-loving element depletions in the lunar mantle can be reconciled with segregation of a 2.5 mass% lunar core (Steenstra et al., 2016). Red line is the liquidus at 5 GPa and yellow vertical bar represents the estimated lunar core S content necessary to sustain a core dynamo for the estimated period (Laneuville et al., 2014) (reproduced from Steenstra et al., 2016).*

**Outlook**

Lunar science is currently amid a revival, driven by a renewed space race with new contenders, including public space programs and private endeavors. As outlined above, China completed its first sample-return mission Chang'e 5, returning the youngest ever sampled lunar basalts with an age of 1963 ± 57 Ma on the 16th of December 2020 (Che et al., 2021). On the 14th of July 2023 India became the fourth nation to successfully land a spacecraft on the surface of the Moon with the mission Chandrayaan-3, at the southernmost landing site on the surface of the Moon to date (69.383°S) (Singh et al., 2023). The latest nation with a successful lunar landing was Japan on the 19th of January 2024 with SLIM (Smart Lander for Investigating Moon), which included two small rovers. NASA is running a two-fold strategy for a return to the Moon. First, Artemis aims to land the first woman on the lunar south pole, initially by 2024 (Smith et al., 2020). The Artemis 2 mission with a lunar flyby is currently scheduled no earlier than September 2025, and a lunar landing with Artemis 3 at the earliest in September 2026. Second, the NASA lunar program aims to integrate commercial lunar payload services. In this strategy private companies deliver science and technology payloads to the lunar surface to catalyze lunar economic growth (NASA, 2020). For example, the Volatiles Investigating Polar Exploration Rover (VIPER) will be delivered by this commercial lunar payload service with a payload aimed at the detection and analysis of volatile elements in permanently shadowed areas in the south pole region, which may include S-bearing deposits (Berezhnoy et al., 2003; Colaprete et al., 2022; Smith et al., 2022). Odysseus by Intuitive Machines became the first privately owned spacecraft to successfully land on the Moon on the 22$^{nd}$ February 2024. In the longer term the Lunar Gateway, which includes a collaboration with ESA, will provide easier access to the lunar surface. In preparation for the return to the Moon the Apollo Next Generation Sample Analysis (ANGSA) program has opened the Apollo 17 double drive tube 73001/73002, with the intent to apply state of the art analytical methods to pristine lunar samples. The drive tube contains basalt fragments and pyroclastic glasses and will provide new insights into the behavior of volatiles including S in lunar volcanism (McCubbin et al., 2022; Shearer et al., 2023; Gross et al., 2023; Yen et al., 2023). In summary, a growing interest in lunar science and volatile elements in particular, with upcoming crewed missions, robotic exploration, and new returned samples will likely reshape our understanding of S on the Moon.

## 3. Mercury

Prior to the NASA MESSENGER mission that entered Mercury's orbit on March 18[th] 2011 (Solomon and Anderson, 2018) very little was known about the chemistry of Mercury and its surface, including the abundance of S (Solomon, 2003). The Mariner 10 probe had previously established that Mercury had a large iron-rich core and only a thin silicate mantle (420 ± 30 km depth, Hauck et al., 2013), large intercrater plains were likely of volcanic origin, lobate scarps were the result of global contraction due to cooling of the lithosphere, and that Mercury experienced a similar intense bombardment in its history as the Moon (Strom, 1979). Sprague et al. (1995) used Earth-based telescopic radar and microwave observations (e.g., NASA/JPL Goldstone 70-m transmitter, and Very Large Array (VLA) at the National Radio Astronomy Observatory) of Mercury to infer a high S abundance on the surface. Specifically, bright radar spots at Mercury's poles were interpreted as evidence for the presence of elemental S in permanently shadowed cold-traps. Both microwave transparency of surface materials and a high index of refraction are evidence for the presence of sulfides, including troilite, pyrrhotite, daubréelite, alabandite, sphalerite, and oldhamite, as components in regolith mixtures with silicate glasses of impact origin, though the $Fe^{2+}$ was likely very low (Sprague et al., 1995; Solomon, 2003). To test the hypothesis of a high S abundance on the surface of Mercury, Sprague et al. (1995) suggested a combination of UV, mid-IR, and microwave spectral observations from Earth-based observatories. The nature of the radar-reflective materials at the poles, as well as the volatile element inventory of Mercury, were among the primary scientific objectives of MESSENGER (Solomon and Anderson, 2018).

### Remote sensing MESSENGER S-observations

MESSENGER was equipped with three sensors for chemical mapping, including the X-ray spectrometer XRS, the γ-ray spectrometer GRS, and the neutron spectrometer NS (Solomon and Anderson, 2018). The XRS and the GRS provided information on S abundances on Mercury's surface. At an energy range of 1–10 keV the XRS detected major rock forming elements Mg, Al, Si, S, Ca, Ti, Cr, Mn, and Fe at the very surface of the planet and only to a depth of few tens of microns (Schlemm et al., 2007; Nittler et al., 2018). The X-ray emission of elements heavier than Si was only sufficiently high during rare solar flares, resulting in gaps in the planetary coverage (Nittler et al., 2018). The elemental abundances measured on the surface were reported as mass ratios to Si (Nittler et al., 2011; Weider et al., 2012, 2014, 2015). Figure 13 shows the global map of the S distribution on the surface of Mercury as a mass ratio of S/Si with a global average of $0.076 \pm 0.019$ (Nittler et al., 2020).

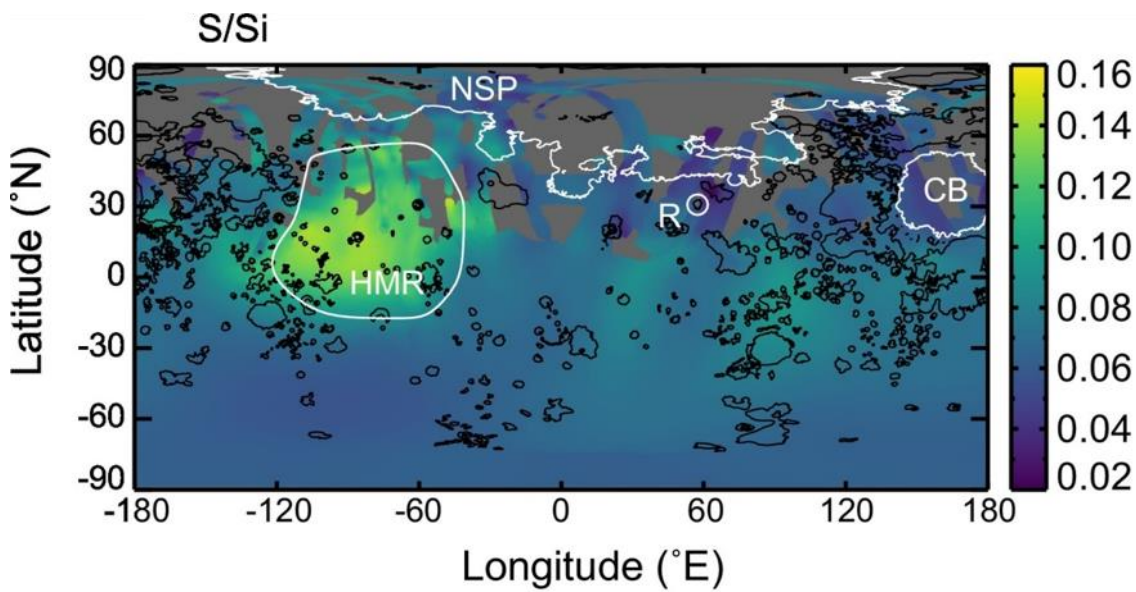

*Figure 13. Sulfur distribution on the surface of Mercury expressed as S/Si (Nittler et al., 2020). The highest S abundances 90 degrees west at the equator and north of it coincides with the highest Ca and Mg abundances, referred to as the high magnesium region (HMR) (reproduced from Nittler et al., 2020).*

The GRS determined abundances of C, O, Na, Al, Si, K, S, Cl, Ca, Fe, Th, and U by measuring γ-ray emitted by the natural decay of radioactive elements, and the interaction of chemical elements with galactic cosmic rays. These γ-rays originate from a depth of up to tens of centimeters (Goldsten et al., 2007). The GRS determined a global surface S/Si mass ratio of $0.092 \pm 0.015$ (Peplowski et al., 2011, 2015, 2016; Evans et al., 2012; Nittler et al., 2018).

Based on the elemental ratios from XRS and GRS data the average compositions of distinct chemical terranes were determined by treating Si as a free parameter (Lawrence et al., 2017; Vander Kaaden et al., 2017; McCoy et al., 2018). The $SiO_2$ concentrations determined by these authors range from 49 to 64 wt.%, in the range of basalts to andesites. The FeO contents fall below 2 wt.%, though it remains unclear if iron occurs mostly as ferrous oxide, sulfide, or metal. The largest variation is observed in the MgO abundance, from 11 to 24 wt.%, informing the distinction of high-Mg and low-Mg chemical terranes (Weider et al., 2015; Lawrence et al., 2017; Vander Kaaden et al., 2017; McCoy et al., 2018; Nittler et al., 2018). The highest S concentration coincides with the high-Mg terrane (Figure 13) with an abundance of ~3 wt.% (Vander Kaaden et al., 2017; McCoy et al., 2018). Notably, the S/Si elemental weight ratio correlates with that of Ca/Si, which was interpreted as evidence for the presence of the CaS mineral oldhamite (Weider et al., 2016).

MESSENGER also carried spectrometers providing some information on the surface mineralogy, the Mercury Dual Imaging System (MDIS) and the Mercury Atmospheric and Surface Composition Spectrometer (MASCS). These spectrometers covered wavelengths from ultraviolet to short-wave infrared (up to 2.5 µm) (Hawkins et al., 2007; McClintock and Lankton, 2007). The main observation is that spectral variation is most evident in changes in the spectral slope, distinguishing high-reflectance red plains and low-reflectance material (Murchie et al., 2015, 2018). Low-reflectance suggests the presence of graphite at the surface of Mercury. In contrast, specific sulfides or ferrous iron bearing minerals could not be identified unambiguously (Murchie et al., 2015). Bright areas with red spectral features were interpreted as the products of pyroclastic deposits (Head et al., 2008, 2009; Kerber et al., 2009, 2011; Prockter et al., 2010; Weider et al., 2015; Murchie et al., 2018). Spectral properties of hollows (irregular shallow depressions, possibly associated with volatile loss (Blewett et al., 2018)) could be related to the occurrence of MgS (niningerite) (Vilas et al., 2016)

Nathair Facula, an asymmetrical bright patch NE of Rachmaninov crater and west of Copland crater, is the largest bright patch on the surface of Mercury and interpreted as the deposit of a pyroclastic eruption (Weider et al., 2016). Importantly, the area shows a significant depletion of S and C (NE Rach in Figure 13). This was interpreted to show loss of S and C during the eruptive displacement of the pyroclastic deposit, and suggests that the explosive volcanic eruption was driven by a S- and C-bearing volcanic gas (Kerber et al., 2011; Nittler et al., 2014; Weider et al., 2016)

## Sulfides on Mercury

The sulfide mineralogy at very reducing conditions below IW-3, relevant for conditions inferred for Mercury, enstatite chondrites, and the aubrite parent body (Keil, 2010), includes the minerals troilite (FeS), Ti-Cr-bearing troilite, daubréelite (FeCr$_2$S$_4$), djerfisherite (K$_6$Na(Fe$^{2+}$,Cu,Ni)$_{25}$S$_{26}$S), caswellsilverite (NaCr$_2$S$_2$), as well as oldhamite (CaS), niningerite (MgS), and alabandite (MnS) (Anzures et al., 2020a; see a detailed review of sulfide mineralogy in Bindi et al., 2024, this volume). The ionic sulfides oldhamite and niningerite are predicted to be the primary sulfides in the source of the Fe-poor Mercury basalts (Namur et al., 2016). Oldhamite in particular is a potential host for REEs on Mercury and may play a similar role to KREEP on the Moon (Crozaz and Lundberg, 1995; Gannoun et al., 2011; Hammouda et al., 2022).

The sulfides oldhamite, niningerite, and alabandite may have a significant effect on the surface reflectance spectra of Mercury in the visible and near-infrared spectral range (Murchie et al., 2018). The sulfides have characteristic absorption features at 0.5 – 0.7 µm that may be detectable by reflectance spectroscopy. Heating of CaS to Mercury daytime temperatures (up to 430 °C) results in a darkening, and for MgS a shift to longer wavelengths (reddening) (Helbert et al., 2013), though these sulfides are overall brighter compared to the average regolith on Mercury (Murchie et al., 2018). However, oldhamite may remain largely obscured in MESSENGER VIRS (Visible and Infrared Spectrograph) data, due to the high signal-to-noise ratio (Izenberg et al., 2014). In the mid-infrared spectral range (7 – 14 µm wavelength) the sulfides have few characteristic spectral absorption properties. Specifically, oldhamite was proposed to have spectral absorption bands near 7.4 and 9.6 µm, and niningerite an absorption band at 7.8 µm (Varatharajan et al., 2019). Olhamite and Niningerite form a solid solution only at high temperature (~1200 °C). Synthetic sulfides of this complete solid solution were investigated by mid-infrared spectroscopy and show no clear spectral features, as expected from the ionic nature of the sulfides (Reitze et al., 2024). Reitze et al. (2024) showed that pure CaS is particularly susceptible to alteration when exposed to air, which may explain the strong spectral features observed by Varatharajan et al. (2019). Nevertheless, spectral properties determined in the laboratory may be used by the MERTIS (Mercury Radiometer and Thermal Imaging Spectrometer) instrument on board the BepiColombo probe, in particular when considered in synthetic multiphase mixtures with minerals and glasses (Varatharajan et al., 2019; Hiesinger et al., 2020; Morlok et al., 2021; Morlok et al., 2023; Renggli et al., 2023b; Reitze et al., 2024).

At very low oxygen fugacities, where $S^{2-}$ replaces $O^{2-}$ on the anion sublattice of glasses and melts in a pseudo-equilibrium to become a major anion (Fincham and Richardson, 1954; O'Neill and Mavrogenes, 2002), melt structures change, affecting the partitioning behavior of major and trace elements, as well as the physical properties of the melts. Specifically, the bonding environment of divalent cations changes and $S^{2-}$ preferentially bonds with $Mg^{2+}$ and $Ca^{2+}$ at conditions more reducing than IW-3, shown in Figure 1 (Anzures et al., 2020b).

The S speciation in experimental glasses measured by XANES as a function of $fO_2$ (Figure 1) suggests that the Fe detected on the surface of Mercury (0.59 – 2.44 wt.% Fe, McCoy et al. 2018) occurs entirely as FeS, instead of FeO. Anzures et al. (2020b) suggests that $fO_2$ models based on the assumption that FeO occurs in the melt may therefore be in error. If FeS and $FeCrS_4$ exsolved from the melt early in the magmatic history, were entrained and transported to the surface, they may account for up to 25% of the observed S on Mercury's surface. The consequence of these surface sulfides would be a reduced S solubility, favoring a more oxidized mantle source closer to IW-3, rather than IW-7 (Anzures et al., 2020b).

The high abundance of S remotely observed on the surface of Mercury was interpreted to be hosted as magmatic sulfides (Zolotov et al., 2013; Namur et al., 2016; Cartier and Wood, 2019). If the S remained in the lavas upon eruption, either dissolved in the melts or crystallized as sulfides, then the high abundance suggests lack of oxidation of the magma in the crust prior to eruption (Namur and Charlier, 2017; Deutsch et al., 2021). On the other hand, assimilation of oxides into the melts, as they ascended to the surface, could have partially oxidized S, allowing the degassing of $S_2$, $CS_2$, and COS gas (Figure 2) (Zolotov, 2011; Cartier and Wood, 2019). These gas species would have been major components in the gas phase driving pyroclastic eruptions, together with CO (Weider et al., 2016). Graphite-melt smelting has been suggested as an alternative mechanism to form CO gas to drive explosive volcanic eruptions on Mercury (McCubbin et al., 2017; Iacovino et al., 2023), similar to a graphite oxidation process releasing CO proposed for lunar pyroclastic eruptions (Fogel and Rutherford, 1995). Experimental smelting of FeO in a silicate melt and graphite above 1100 °C released a CO-rich gas with minor S-abundances (McCubbin et al., 2017; Iacovino et al., 2023):

$$yC_{graphite} + M_xO_{y,melt} \rightarrow yCO_{gas} + xM^0$$

Where M is a metal, x and y denote the number of metal and oxygen atoms in the smelting reaction. Iacovino et al., (2023) propose that all except the largest pyroclastic deposits on Mercury could have formed by a combination of such a smelting reaction and additional S-H-degassing.

High-temperature S-rich gases are very reactive when they come in contact with silicate glasses, melts and minerals, forming sulfates at oxidizing conditions, and sulfides at reducing conditions (Renggli and King, 2018; Renggli et al., 2019b). Experiments on the reaction of a reduced S-rich gas (Figure 2) with synthetic Mercury glasses result in the formation of oldhamite, niningerite, and Fe-Ti-sulfides (Renggli et al., 2022). The reaction of $S_2$ gas with a Mg-, Ca-, and Fe-bearing aluminosilicate glass or melt in the presence of graphite (Peplowski et al., 2016) can be written as:

$$\{MgO + CaO + FeO + SiO_2 + Al_2O_3\}_{glass;\,melt} + 1.5S_{2(g)} + 3C$$
$$\rightarrow \{SiO_2 + Al_2O_3\}_{glass;\,melt;\,mineral} + 3(Mg,Ca,Fe)S + 3CO_{(g)}$$

This sulfidation reaction is experimentally supported for conditions on Mercury and may result in the observed enrichment of S on the planet's surface (Renggli et al., 2022). Like glasses or melts, silicate minerals such as anorthite, forsterite, or diopside react with a reduced S-rich gas to form oldhamite and niningerite (Renggli et al., 2023b). Figure 14 shows Gibbs free energy minimization calculations for anorthite, forsterite, and diopside in a S-rich reduced gas at 1000 °C and 1 bar pressure, and the stability of the sulfides at reduced Mercury conditions (Renggli et al., 2022). For example, the sulfidation of forsterite follows the reaction equation:

$$Mg_2SiO_4 + S_{2(g)} + 2C \rightarrow SiO_2 + 2MgS + 2CO_{(g)}$$

This formation of niningerite by silicate sulfidation has also been suggested in EH3 enstatite chondrites (Lehner et al., 2013). The result of the forsterite sulfidation reaction is shown in Figure 15 (Renggli et al., 2023b). The sulfide coating consists of niningerite grains with sizes of up to 10 µm. The degree of sulfidation is kinetically limited by the diffusion of the sulfide forming cation (e.g. Mg, Ca, Fe) from the interior of the silicate mineral or glass to the surface where the reaction occurs (Renggli and King, 2018), and as the sulfide coating grows, the reaction rate decreases and becomes self-inhibiting (Renggli et al., 2019a). The increase in the degree of sulfidation with increasing temperature is apparent in the case of forsterite sulfidation (Figure 15), where the second solid reaction product ($SiO_2$) is clearly identified in the sulfidation experiment at 1200 °C over 24 hours. Except for the northern volcanic plains, quartz is not predicted as a major rock-forming mineral in the volcanogenic lithologies on the surface of Mercury (Vander Kaaden and McCubbin, 2016; Namur and Charlier, 2017). Elevated abundances of quartz associated with niningerite and oldhamite could therefore be interpreted as evidence for sulfidation reactions and secondary S enrichment on the surface of Mercury (Renggli et al., 2022; Renggli et al., 2023b). Consequently, if S was secondarily enriched on the surface of Mercury by sulfidation, S solubilities in the Mercury magmas would be lower and the oxygen fugacity closer to the more oxidized end of the estimated range (IW-7 to IW-3).

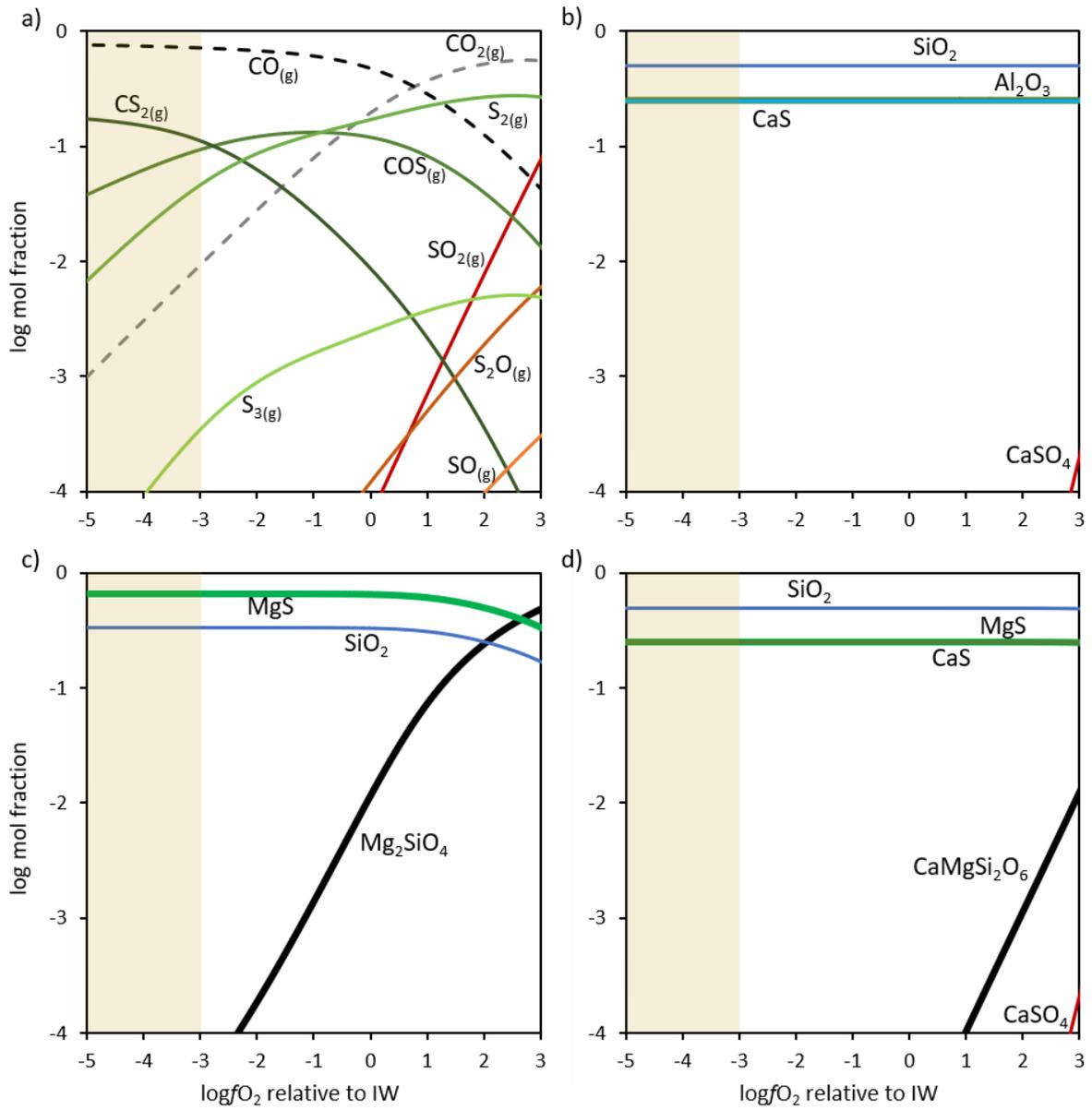

*Figure 14. Gibbs free energy minimization calculations of equilibria between a S-rich gas at 1000 °C and 1 bar (a), as a function of oxygen fugacity with (b) anorthite, (c) forsterite, and (d) diopside. The mol fractions of the solids (b–d) are normalized to the total amounts of solids in each model. The shaded area indicates reduced conditions below IW-3, proposed for magmas on Mercury. At these conditions the model predicts the stability of oldhamite (CaS) and niningerite (MgS) (reproduced from Renggli et al., 2022).*

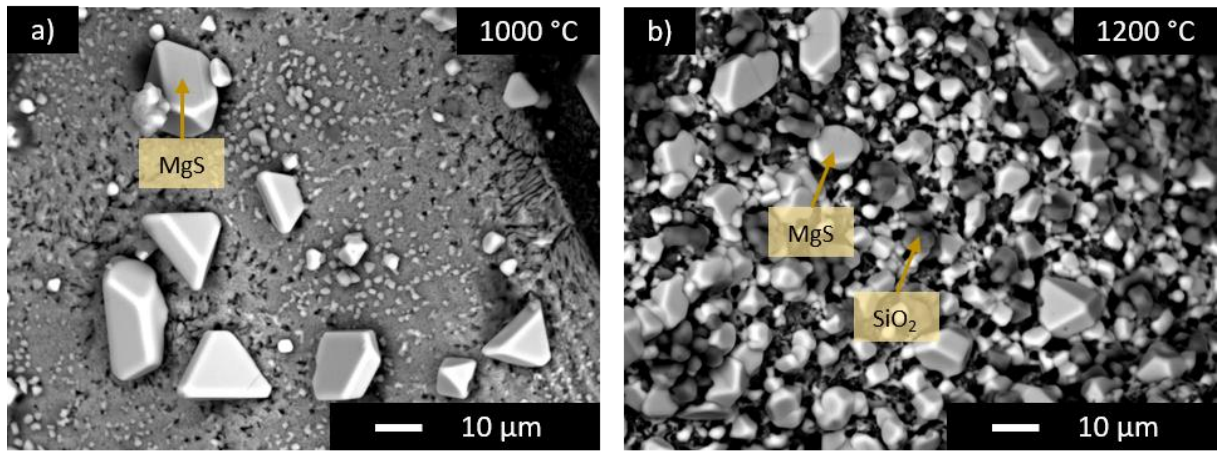

*Figure 15. Back-scattered electron images of the surfaces of forsterite crystals reacted in an evacuated and sealed silica glass ampule with a reduced S-gas. The images show the formation of niningerite on the crystal surfaces after 24h reaction at a) 1000 °C and b) 1200 °C. With increasing temperature and experimental duration, the degree of reaction increases. At 1000 °C individual MgS grains are observed on the surface, and at 1200 °C the reaction has further progressed and both MgS and $SiO_2$ grains can be identified on the forsterite crystal (Renggli et al., 2023b).*

Mercury hollows are a unique type of landform, and their origin has been related to the decomposition of volatile bearing minerals, including sulfides. The hollows are depressions (24 m deep on average) with irregular shapes, diffuse borders, and elevated reflectance (Blewett et al., 2013, 2018; Thomas et al., 2014). The origin of the formation of these depressions remains enigmatic. The irregular halos surrounding the depressions have been interpreted as evidence for the mobilization of a volatile phase and may be the consequence of recent secondary remobilization of sulfides (Blewett et al., 2011, 2016; Vaughan et al., 2012; Helbert et al., 2013; Vilas et al., 2016; Lucchetti et al., 2018; Varatharajan et al., 2019; Barraud et al., 2020, 2023; Wang et al., 2020; Phillips et al., 2021; Renggli et al., 2022). The formation of the hollows may also be related to the intense space weathering on Mercury, resulting in the decomposition of mineral phases that contain volatiles. This has been investigated experimentally by photon-stimulated desorption of the sulfides oldhamite (Bennett et al., 2016) and niningerite (Schaible et al., 2020). Ionized atomic $S^+$ was observed by the Fast Ion Plasma Spectrometer (FIPS) on MESSENGER and will be investigated by the PHEBUS (Probing of Hermean Exosphere by Ultraviolet Spectroscopy) and SERENA (Search for Exospheric Refilling and Emitted Natural Abundances) instruments onboard BepiColombo (Benkhoff et al., 2010). Both $S^+$, as well as $Ca^0$ and $Ca^+$, which have also been detected in Mercury's exosphere, have been suggested to originate from sulfide decomposition by intense space weathering, which could explain the formation of the hollows (Bennett et al., 2016; Schaible et al., 2020). Finally, a connection between an elevated abundance of CaS, MgS, and $Na_2S$, and the formation of hollows is supported by near-ultraviolet to near-infrared spectral observations by MESSENGER, with spectral modeling suggesting an enrichment of sulfides in hollows up to two times higher compared to Mercury's high-reflectance smooth plains, where hollows are rare (Barraud et al., 2023).

### S in Mercury's interior

Mercury's building blocks were likely highly reduced and metal-rich, resulting in the formation of a large Fe-rich, Si-bearing metallic core, though Mercury may have lost a large part of its silicate mantle in a large-scale impact event (Ebel and Stewart, 2018). Sulfur is incompatible in Si-bearing Fe liquid, and the incompatibility increases with increased Si contents (Kilburn and Wood, 1997). Due to the overall incompatibility of S in Fe-Si liquid, it is expected that virtually all S would be preferentially partitioned into the silicate magma ocean of Mercury, instead of the core (Steenstra and van Westrenen, 2020). The incompatibility of S in the core, in conjunction with the high bulk S contents of the most plausible meteoritic building blocks inferred for Mercury (i.e. enstatite chondrites), is expected to result in a significant enrichment of S in Mercury's minor mantle. This has led some workers to suggest that FeS, or other sulfide species, may have segregated from Mercury's magma ocean when S contents of the magma ocean exceeded the S solubility of that magma ocean melt (Boukaré et al., 2019). It is hypothesized that this could have resulted in the formation of a FeS layer, that would have sunk into the deeper interior and subsequently enveloped Mercury's Fe-Si core (Malavergne et al., 2010; Chabot et al., 2014), either in a solid or liquid state. The presence of such a layer would be consistent with the hypothesis of a dense layer overlying a dynamo-generating Mercurian Fe-Si core (Hauck et al., 2013). Electrical measurements on sulfide samples at high pressure of up to 10 GPa suggest that a FeS layer at the core-mantle boundary should be liquid and thermally insulating, affecting the long-term stability of the magnetic field (Pommier et al., 2019).

However, experimental and meteoritic studies have shown that significant S can be dissolved in highly reduced, i.e. low FeO, melts. For example, low FeO basaltic vitrophyres in enstatite achondrites (aubrites), have S contents of up to 2.6 wt.% (Fogel, 2005). Experimental studies have confirmed such high S solubilities for low-FeO (O'Neill and Mavrogenes, 2002; Wykes et al., 2015; Wood and Kiseeva, 2015; Namur et al., 2016; Wohlers and Wood, 2017; Steenstra et al., 2020c, b; Anzures et al., 2020a, b). Sulfur solubilities vary dramatically within the low FeO regime with only slight variations in FeO contents (see e.g. Fig. 3 in Steenstra et al., 2020a). Considering the overall uncertainties related to the redox state of Mercury's mantle from surface chemistry, it is unclear whether sulfide liquid segregation occurred during Mercury's differentiation. Cartier et al. (2020) used the Ti/Al measured for Mercury's surface, and the partitioning behavior of Ti at highly reduced conditions, to argue against the presence of a FeS melt layer in Mercury's deep interior (see also Boukaré et al., 2019). Models for the Mercury core further suggest that a hypothetical molten FeS layer above the core would have only minimal effect on the long-term thermal and magmatic evolution of Mercury (Davies et al., 2024). Besides FeS, likely potential candidates for sulfides in Mercury's mantle are CaS and MgS (up to 20 wt.%, e.g., Lark et al. (2022) and references therein), which are much less dense than the surrounding mantle and would constitute a major low-density reservoir. The presence of such a low-density reservoir in Mercury's interior would thus require a denser and/or larger core to satisfy Mercury's bulk density. Finally, the abundances of these sulfide phases again critically depend on the bulk S composition of Mercury as well as potential loss of S during accretion of Mercury.

### BepiColombo mission to Mercury

The NASA MESSENGER mission identified the high abundance of S on the surface of Mercury. As discussed above, the observation has sparked a lively debate about the origin of the high S concentration, the mineralogy of S-bearing phases on the surface, and processes from the planet's origin, through a magma ocean phase, partial melting of the mantle, to pyroclastic eruptions, in which S has played an important role. The dual-spacecraft BepiColombo, consisting of a Mercury Planetary Orbiter (MPO) from the European Space Agency (ESA), and a Mercury Magnetospheric Orbiter from the Japanese Aerospace Exploration Agency (JAXA), will insert into a Mercury orbit in December 2025 (Benkhoff et al., 2021). The MPO includes payload that will further constrain the chemistry of the surface of Mercury, as well as the mineralogy, including areas of the planet's surface that have not been mapped in detail by MESSENGER (Benkhoff et al., 2010, 2021; Rothery et al., 2010). Four instruments on the MPO will investigate the elemental and mineralogical surface composition of Mercury. The Mercury Gamma-ray and Neutron Spectrometer MGNS (Mitrofanov et al., 2021) and the Mercury Imaging X-ray Spectrometer MIXS (Fraser et al., 2010) will measure the surface chemistry. The MGNS has a spatial resolution of 400 km and will measure the composition of the top $1 - 2$ meters for major elements Si, O, C, Al, Fe, Na, and Ca, as well as volatiles S, H, and Cl, and radioactive isotopes including K, Th, and U (Mitrofanov et al., 2021). The MIXS will measure fluorescent X-ray emissions for Si, Ti, Al, Fe, Mg, Na, Ca, P, Mn, K, S, Cr, Ni, and O, over the energy range of $0.5 - 7.5$ keV, that are induced by solar X-ray, solar wind electrons, and protons, to a depth of approximately 10 cm. The spatial resolution of MIXS exceeds that of MGNS with a first channel (collimated MIXS-C) measuring at a resolution of $70 - 270$ km, and a second, imaging X-ray telescope channel (MIXS-T) allowing measurements during major solar flares to resolve features at a scale of less than 10 km (Fraser et al., 2010). This is of particular importance to resolve variations in S abundances in pyroclastic deposits, hollow-forming regions, and polar volatile deposits. The chemical measurements will be combined with spectral observations from the visible to the mid-infrared. The SIMIO-SYS (Spectrometer and Imaging for MPO BepiColombo Integrated Observatory SYStem), composed of a high-resolution camera (HRIC), a stereo camera (STC), and a visible and near-infrared hyperspectral imager (VIHI), and will map the entire surface of the planet at a maximum spatial resolution of $5 - 12$ m/px. The VIHI will map the entire surface in the spectral range of $400 - 2000$ nm at a spatial resolution of 480 m/px, and has a 20% high resolution coverage at 120 m/px (Cremonese et al., 2020). The Mercury Radiometer and Thermal Infrared Spectrometer (MERTIS) will operate in the spectral ranges of $7 - 14$ µm (spectrometer) and $7 - 70$ µm (radiometer), with a spectral resolution of 90 nm and at a spatial resolution of 500 m or better, for a global mapping coverage. In the mid-infrared spectral range MERTIS is specifically sensitive to the identification of the silicate mineralogy on the surface of Mercury (Hiesinger et al., 2020). The spectral observations will also provide constraints on the role of sulfides in hollow-forming processes, as well as the association of S with other potential volatile components and graphite-rich regions (Thomas et al., 2016; Barraud et al., 2023). PHEBUS (Probing of Hermean Exosphere By Ultraviolet Spectroscopy) will investigate the composition and structure of the Mercury exosphere and magnetosphere, as well as the physical and chemical coupling with the surface of the planet (Quémerais et al., 2020). This will build on MESSENGER observations with the Mercury Atmospheric and Surface Composition Spectrometer (MASCS), which detected H, He, Na, K, Ca, Mg, Al, Fe, and Mn as exospheric neutral species (McClintock et al., 2018). PHEBUS specifically aims to detect new species including atomic S at a far ultra-violet wavelength of 180.7 nm, in addition to other atomic, molecular, and ionic species (Quémerais et al., 2020). Finally, the SERENA instrument suite (Search for Exospheric Refilling and Emitted Natural Abundances) will measure neutral particles in the exosphere, charged particles in the magnetosphere, and the solar wind (Orsini et al., 2021). These observations will be key to understand the role of S and other volatiles in hollow formation, accumulation in permanently shadowed regions, and the interaction of surface processes with the exosphere and magnetosphere of Mercury (Bennett et al., 2016; Schaible et al., 2020).

## Acknowledgments

Renggli acknowledges the funding from the Deutsche Forschungsgemeinschaft (DFG, German Research Foundation – project 442083018. Steenstra was funded through a Marie Skłodowska-Curie Postdoctoral Fellowship (101020611).